\documentclass[preprint,12pt]{elsarticle}

\graphicspath{{./}}

\usepackage{graphicx}
\usepackage{amsmath}
\usepackage{amssymb}
\usepackage{bm}
\usepackage{booktabs}
\usepackage{makecell}
\usepackage{subcaption}
\usepackage{textcomp}
\usepackage{xcolor}
\usepackage{enumitem}
\setlist[enumerate]{itemsep=2pt, parsep=1pt, topsep=3pt}
\usepackage{hyperref}   

\biboptions{numbers,sort&compress}

\newcommand{\Cpp}{C\nolinebreak\hspace{-.05em}\raisebox{.4ex}{\tiny\bf +}%
                  \nolinebreak\hspace{-.10em}\raisebox{.4ex}{\tiny\bf +}}

\journal{Computer Physics Communications}

\makeatletter
\def\ps@pprintTitle{%
  \let\@oddhead\@empty
  \let\@evenhead\@empty
  \def\@oddfoot{\footnotesize\itshape
     Preprint submitted to Computer Physics Communications
     \hfill\today}%
  \let\@evenfoot\@oddfoot}
\makeatother

\begin{document}

\begin{frontmatter}

\title{IteraSim RAG: A Multi-Stage Retrieval-Augmented Agentic
       Back-End for OpenFOAM-Based Computational Fluid Dynamics}

\author[eth]{Pratyush Kumar\corref{cor1}}
\ead{pratyush.ethz@gmail.com}
\cortext[cor1]{Corresponding author.}

\address[eth]{Seminar for Applied Mathematics, ETH Z\"urich,
              R\"amistrasse 101, 8092 Z\"urich, Switzerland}

\begin{abstract}
Configuring a computational fluid dynamics (CFD) case in
OpenFOAM requires assembling a multi-directory input deck of
mutually consistent solver, discretisation and
boundary-condition dictionaries -- a task that remains a
substantial barrier to non-specialist use of open-source CFD
software. Large language models (LLMs) coupled with
retrieval-augmented generation (RAG) can lower this barrier,
but existing systems retrieve with a single flat query, apply
one retrieval strategy to operationally distinct requests, and
let a single agent both draft and review its own output.
We present \textbf{IteraSim RAG}, a retrieval-augmented
software back-end for automated OpenFOAM case generation built
around these three limitations.
Retrieval proceeds in three stages.
An LLM first expands the query into physics, solver-keyword and
troubleshooting variants, Reciprocal Rank Fusion then merges
the resulting ranked lists, and Maximal Marginal Relevance
re-ranks the fused candidates against an HNSW-indexed dense
vector store.
A deterministic keyword router dispatches tool-conditioned
workflow queries and corpus-wide physics queries down separate
retrieval paths, and generation is split across an Architect,
an InputWriter and a Reviewer agent, backed by a static
canonical-knowledge layer covering solver selection, turbulence
closures, boundary conditions and finite-volume defaults.
On an openly released 28-case benchmark spanning zero-shot
setup, few-shot generalisation, single-parameter modifications
and turbulence-model swaps, the pipeline attains a mean
retrieval coverage of $77.9\%$ (median $79.1\%$), with the
parameter-modification category exceeding $90\%$.
All six reference configurations run to completion on
OpenFOAM~v2506, and two synthetically corrupted cases are
diagnosed and repaired within the bounded Reviewer loop using
only the solver log and the canonical layer.
The benchmark specification, scoring rubric, corrupted-case
probes and figure-regeneration scripts are released for
reproducibility.
\end{abstract}

\begin{keyword}
Retrieval-augmented generation \sep
Large language models \sep
Multi-agent systems \sep
Computational fluid dynamics \sep
OpenFOAM \sep
Scientific software automation
\end{keyword}

\end{frontmatter}

\thispagestyle{pprintTitle}


\section{\label{sec:intro}Introduction}

Computational fluid dynamics (CFD) underpins a broad range of
engineering and scientific applications, spanning laminar
transport in pipe networks~\cite{Ismail2012},
centrifugal turbomachinery~\cite{Alemi2015},
membrane-based gas separation~\cite{Safaei2021},
haemodynamic transport through branching
vasculature~\cite{Sochi2015},
coupled CFD--DEM analysis of packed
beds~\cite{Kumar2023a}, and helium cooling of pebble-bed
nuclear reactor canisters~\cite{Kumar2021}, among many others.
Within open-source CFD tools, OpenFOAM is the most widely used
software because its object-oriented \Cpp{} architecture allows
arbitrary modification of solvers, discretisation schemes and
physical sub-models without the licensing constraints of
commercial packages~\cite{openfoam,Weller1998}.
OpenFOAM has been applied across a comparably broad range of
physical regimes, including wall-bounded and free-shear
turbulence, incompressible and compressible flows, multiphase
and free-surface transport, conjugate and buoyancy-driven
heat transfer, porous-media and granular flows, and reactive
and combustion modelling~\cite{Samal2023,Kumar2025,%
Kumar2021,Kumar2023a}.
That flexibility carries a steep set-up burden.
Configuring a typical case means assembling a multi-directory
folder structure, picking discretisation and linear-solver
settings that are mutually compatible, writing physically
consistent boundary conditions, and then diagnosing the
divergence that a poor combination of any of these will
trigger.
Adding new physics to an existing solver is harder still, even
for seasoned users.
The associated expertise is acquired over years of practice and
remains the principal obstacle to wider adoption of OpenFOAM
outside the dedicated CFD community~\cite{Pandey2025,Yue2025}.

A complementary line of work has therefore explored data-driven
assistance for CFD users.
Early efforts focused on surrogate modelling of specific
physics, including Reynolds-averaged turbulence
closures~\cite{Duraisamy2019}, subgrid-scale stresses for
large-eddy simulation~\cite{Beck2019}, and reduced-order
models for parametric design~\cite{Brunton2020}.
Large language models (LLMs) have changed what such an
assistant can be.
Instead of executing fixed code paths, an LLM-based tool can
read a problem stated in plain language, draft its own
implementation plan, and revise that plan as the solver output
comes back.
In materials science, agentic systems based on LLMs have been
shown to autonomously propose and refine candidate
alloys~\cite{Boiko2023}; in scientific software more broadly, they
have demonstrated the ability to write, debug and execute
non-trivial code in closed loops~\cite{Wang2024Agents}.
The application of LLMs to CFD is now a distinct and rapidly
growing sub-field.
Our literature survey identifies the principal LLM-for-CFD
systems and their contributions, summarised in
Table~\ref{tab:prior-vs-present}; each system is discussed in
turn in the paragraphs below.

\begin{table*}[t]
\centering
\caption{Design choices comparison across RAG systems for CFD.}
\label{tab:prior-vs-present}
\setlength{\tabcolsep}{6pt}
\resizebox{\linewidth}{!}{%
\begin{tabular}{l c c c}
\toprule
System &
\makecell{Retrieval} &
\makecell{Drafter/\\Reviewer split} &
\makecell{Loop\\bound} \\
\midrule
OpenFOAMGPT~\cite{Pandey2025}     & Flat        & Single agent  & open \\
MetaOpenFOAM~\cite{Chen2024Meta}  & Flat        & Yes (3 roles) & --   \\
Foam-Agent 1.0~\cite{Yue2025}     & Flat        & Yes           & 25   \\
Foam-Agent 2.0~\cite{Yue2025v2}   & Hier.\ multi-index & Yes    & 25   \\
FoamGPT~\cite{Yue2025Foamgpt}     & Fine-tuned LLM (no RAG) & Planner+writer & -- \\
ChatCFD~\cite{Fan2026}            & Flat + error locator & Yes  & --   \\
CFDAgent~\cite{Xu2025}            & Flat        & Yes           & --   \\
SwarmFoam~\cite{Yang2025}         & Flat        & Swarm (peer agents) & -- \\
\midrule
IteraSim RAG &
Multi-stage &
Architect / InputWriter / Reviewer &
10 \\
\bottomrule
\end{tabular}%
}
\end{table*}

Pandey~\textit{et~al.}~\cite{Pandey2025} introduced
\textit{OpenFOAMGPT}, the first retrieval-augmented LLM agent
reported in the CFD literature.
They showed that a general-purpose foundation model coupled
with a flat vector store can configure and debug single-phase
and multiphase OpenFOAM cases through an iterative correction
loop.
Chen~\textit{et~al.}~\cite{Chen2024Meta} introduced
multi-agent role separation with \textit{MetaOpenFOAM}, in
which a planner, input-writer and reviewer collaborate over a
LangChain-based RAG built from the official tutorials,
reporting an 85\% pass rate on an eight-case benchmark.
Yue~\textit{et~al.} extended the multi-agent paradigm to the
full simulation lifecycle in \textit{Foam-Agent}~1.0
and~2.0~\cite{Yue2025,Yue2025v2}, adding a hierarchical
multi-index RAG, dependency-aware generation across coupled
configuration files, a Gmsh-based Meshing Agent and an
MCP-compatible service layer.
The system reaches an 88.2\% end-to-end success rate on a
110-case benchmark.
The same group demonstrated with
\textit{FoamGPT}~\cite{Yue2025Foamgpt} that supervised
fine-tuning on the 202 canonical OpenFOAM tutorial cases can
outperform much larger proprietary baselines.
Here the quality of the training data mattered more than the
size of the model.
\textit{ChatCFD}~\cite{Fan2026} pushed the executability
metric to 82.1\% over 315 cases and introduced a Physics
Interpreter that evaluates whether a runnable case is also
physically meaningful---the first formal treatment of physical
fidelity in this line of work.
\textit{CFDAgent}~\cite{Xu2025} and
\textit{SwarmFoam}~\cite{Yang2025} further explored zero-shot
language-guided orchestration and swarm-based cooperation
across parallel agent instances, respectively.
Across these systems the dominant innovations have been in
agent decomposition (role specialisation, dependency
awareness), in dataset curation (FoamGPT) and in evaluation
(physical-fidelity scoring in ChatCFD).
The retrieval layer itself has remained largely unchanged; a
single user query is embedded and the top-$K$ nearest chunks
are concatenated into the prompt.
Flat, single-query retrieval introduces at least three persistent failure
modes in the systems above.

\paragraph{(L1) Single-query, flat retrieval.}
The user's natural-language query is embedded once and matched
against the chunked corpus by cosine similarity.
This pattern is sensitive to the long-documented vocabulary
mismatch between user phrasing and corpus
terminology~\cite{Furnas1987}.
It also provides no mechanism to suppress near-duplicate
passages, which consume prompt budget without adding new
evidence.

\paragraph{(L2) Undifferentiated routing across query types.}
A request to \textit{``write a \texttt{controlDict} for a
steady \texttt{simpleFoam} channel''} is operationally
distinct from a question such as \textit{``which low-Re
$k$--$\omega$ variant should I use under an adverse pressure
gradient''}.
Yet both are answered with the same flat retrieval over the
same corpus.
The first call wants tightly tool-conditioned excerpts of
documentation; the second wants breadth across the physical
literature.

\paragraph{(L3) Drafting and review entangled in one agent;
no canonical reference layer.}
Even where role-specialised agents are used, the drafter and
the reviewer often share prompts and context, which is known
to amplify hallucinations and inflate the cost of recoverable
errors~\cite{Madaan2023}.
Furthermore, agents have no explicit practitioner-validated
reference for solver syntax, boundary-condition selectors or
discretisation defaults---all details that a foundation model
can easily mis-recall when operating zero-shot over a purely
retrieved knowledge base.

IteraSim RAG is designed to close each of these gaps.
Two further axes not introduced in the previous works
(Table~\ref{tab:prior-vs-present}) complete the comparison.
Prior systems do not implement an explicit dual-mode
retrieval router that separates solver-conditioned workflow
queries from corpus-wide physics-level queries~(L2).
Similarly, only ChatCFD~\cite{Fan2026} augments retrieval with
a static practitioner-knowledge layer -- its Solver Template
Database -- and no other system exposes such a reference
alongside the retrieved context~(L3).

Therefore, this paper introduces \textbf{IteraSim RAG}, an
OpenFOAM-oriented back-end designed around (L1)--(L3).
The overall architecture is summarised in
Fig.~\ref{fig:rag-architecture} and described in detail in
Sec.~\ref{sec:methodology}.
Its main contributions are:

\begin{enumerate}
  \item A three-stage retrieval pipeline that responds to
        (L1). An LLM rewrites the raw query into three
        semantic reformulations spanning a physics facet, a
        solver-keyword facet, and a troubleshooting facet.
        Ranked lists returned by each reformulation and by
        the original query are merged through Reciprocal
        Rank Fusion (RRF)~\cite{Cormack2009}, and the fused
        candidate set is re-ranked by Maximal Marginal
        Relevance (MMR)~\cite{Carbonell1998} to produce a
        diverse, low-redundancy context block. Retrieval is
        performed against an HNSW-indexed vector
        store~\cite{Malkov2020} of dense embeddings, and the
        pipeline does not depend on a particular embedding
        family or vector-store implementation.
  \item A deterministic dual-mode router that responds to
        (L2). Each incoming query is dispatched by a
        keyword-driven classifier into one of two modes.
        \textit{Workflow} mode locks retrieval onto a solver-
        and tool-conditioned sub-corpus and walks the
        upstream/downstream tool-dependency graph.
        \textit{Expert} mode broadens retrieval over the full
        corpus, down-weights verbatim full-case listings and
        applies a more diversity-favouring MMR trade-off.
        The classifier is defined over an eighteen-category CFD
        ontology, so no additional LLM call is required to
        route the query.
  \item An Architect--InputWriter--Reviewer triad,
        supplemented by a canonical-knowledge layer, that
        responds to (L3). The Architect decomposes the
        request into a file manifest and picks a reference
        solver and template.
        The InputWriter then emits the individual OpenFOAM
        dictionaries.
        The Reviewer compiles the case, executes it, parses
        solver and \texttt{wmake} logs, and iterates
        corrective patches up to a bounded depth of ten
        cycles.
        The canonical layer is injected alongside the
        retrieved context.
        It encodes solver-selection rules, turbulence-closure
        decision trees, boundary-condition selectors,
        finite-volume defaults and meshing recipes, so that
        established practitioner knowledge is not lost to
        retrieval noise.
  \item A solver-agnostic and tool-aware extensibility
        layer. The retrieval, routing and orchestration
        components are deliberately decoupled from any
        specific solver. The same pipeline currently serves
        OpenFOAM, \texttt{snappyHexMesh}, \texttt{blockMesh},
        Gmsh, \texttt{pyFoam} and ParaView, and accommodates
        new text-documented solvers by extending the
        ingestion manifest and the tool-dependency graph
        alone.
\end{enumerate}

Section~\ref{sec:methodology} describes the system components;
Sec.~\ref{sec:results} evaluates it on a 28-case benchmark
and examines residual failure modes; and
Sec.~\ref{sec:conclusion} draws conclusions.


\section{\label{sec:methodology}Methodology}

\subsection{\label{subsec:overview}System Overview}

\begin{figure}[t]
  \centering
  \includegraphics[width=0.95\linewidth]{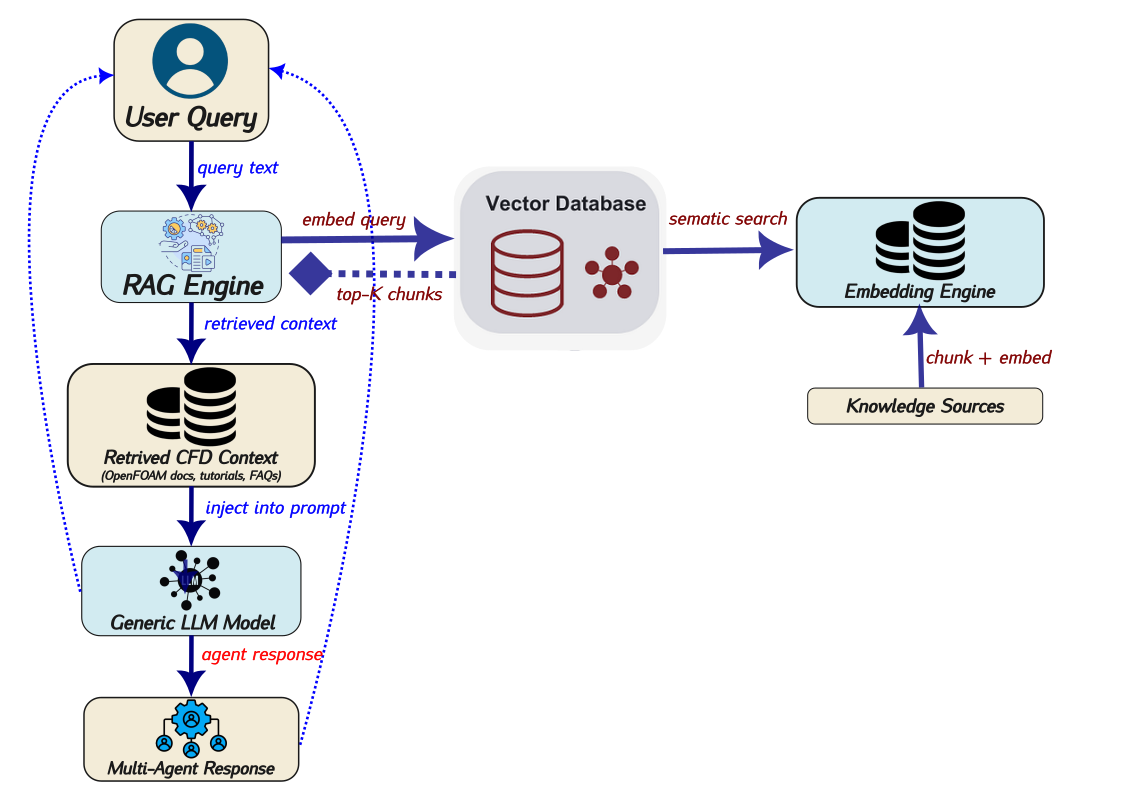}
  \caption{Architecture of IteraSim RAG (each block is walked
  through, top to bottom, in the paragraphs that follow).}
  \label{fig:rag-architecture}
\end{figure}

The overall data flow of IteraSim RAG is summarised
schematically in Fig.~\ref{fig:rag-architecture}.
The figure can be read top-to-bottom as a single pass: a user
query enters at the top, is answered from a retrieval store on
the right, and leaves as a validated OpenFOAM case at the
bottom.
The remainder of this section follows the same path, block by
block; each block in the figure corresponds to a subsection below.

A user query (\emph{User Query} in
Fig.~\ref{fig:rag-architecture}) enters through a
deterministic intent classifier inside the retrieval engine
(\emph{RAG Engine}), which routes it to one of two retrieval
modes (Sec.~\ref{subsec:dual}).
The engine then runs a three-stage retrieval over a vector
store (\emph{Vector Database}) that is populated offline by an
embedding engine (\emph{Embedding Engine}) from the knowledge
sources (\emph{Knowledge Sources}); the corpus and its
ingestion are described in Sec.~\ref{subsec:corpus} and the
retrieval stages in Sec.~\ref{subsec:rag}.
The output is the retrieved CFD context block (\emph{Retrieved
CFD Context}).

This context block, together with a static
canonical-knowledge layer (Sec.~\ref{subsec:canon}), is
injected into the generation stage (\emph{Generic LLM Model}),
realised here as an Architect--InputWriter--Reviewer agent
triad (Sec.~\ref{subsec:agents}).
The triad produces the multi-agent response (\emph{Multi-Agent
Response}); a Runner component executes the case on disk and
returns solver and \texttt{wmake} logs to the Reviewer for
corrective iteration, closing the loop shown by the return
arrows in Fig.~\ref{fig:rag-architecture}.
Implementation and configuration details common to all blocks,
including the provider-agnostic model adapters and the offline
vector-store fallback, are collected in Sec.~\ref{subsec:impl}.

\subsection{\label{subsec:corpus}Knowledge Corpus and Ingestion}

The knowledge corpus (the \emph{Knowledge Sources} of
Fig.~\ref{fig:rag-architecture}) is assembled from five
complementary sources:
(i)~a curated set of expert markdown notes covering solver
selection, boundary-condition idioms, mesh-quality criteria and
numerical-stability heuristics; the full text of these notes is
proprietary, but their section-level structure is documented in
the released evaluation kit~\cite{IteraSimCurated};
(ii)~a troubleshooting frequently-asked-questions set generated
from validated practitioner forum
threads~\cite{CFDOnline} and stored as question--answer pairs;
(iii)~the full local knowledge base of OpenFOAM tutorial cases
and reference dictionaries (more than one thousand markdown
documents) drawn from the official ESI-OpenCFD OpenFOAM
distribution~\cite{OpenFOAMTutorials};
(iv)~the publicly released FoamGPT question--answer pairs
covering the 202 canonical tutorial cases~\cite{Yue2025Foamgpt};
and
(v)~the wave-energy converter (WEC) OpenFOAM extension corpus
released by Sandia National Laboratories~\cite{SandiaWEC},
which supplies coupled multiphase and rigid-body documentation
absent from the upstream tutorials.

Documents are split with a section-aware chunker that respects
markdown level-2 and level-3 heading boundaries.
When no headings are present, it falls back to a
sliding-window word-based splitter.
The default chunk size at ingestion is 400 words for
documentation and 800 words for long-form practitioner notes.
A 100-word overlap preserves semantic continuity across chunk
boundaries and limits the truncation artefacts typical of hard
splits.
Each chunk is then encoded by a general-purpose dense text
encoder into a fixed-dimensional vector ($d{=}3072$ in the
configuration evaluated here).
The vectors are written to the vector store together with
metadata (source URL, tool tag, intent tag, document type and
an MD5 fingerprint used for deduplication) under cosine
distance with the HNSW graph index~\cite{Malkov2020} described
above.
Ingestion is idempotent.
Documents already in the collection (matched by MD5 fingerprint)
are skipped, making ingestion idempotent.

\subsection{\label{subsec:rag}Three-Stage Retrieval Pipeline}

\begin{figure*}[t]
  \centering
  \includegraphics[width=\textwidth]{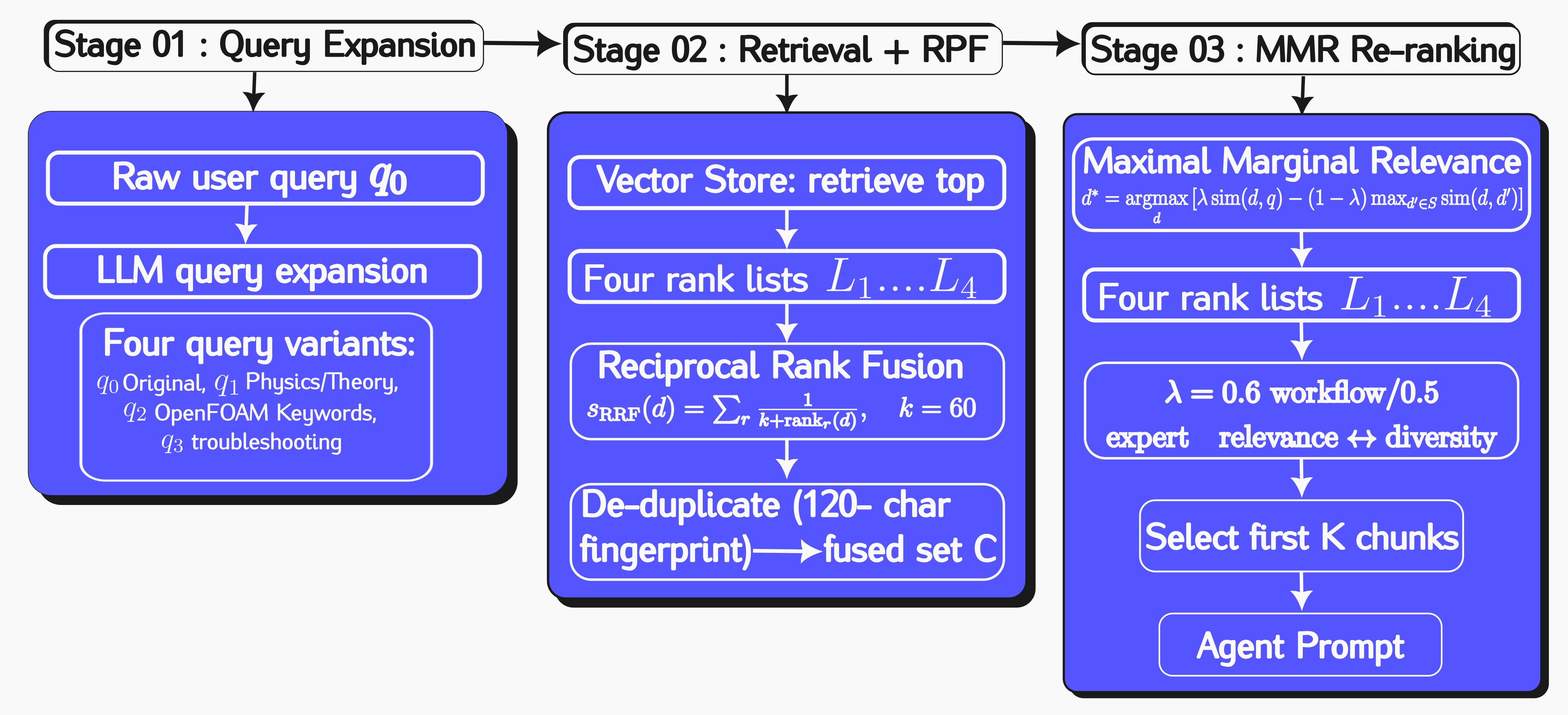}
  \caption{Schematic representation of the three-stage retrieval
  pipeline of IteraSim RAG.}
  \label{fig:rag-pipeline}
\end{figure*}

For every user turn the retrieval layer transforms the raw
query into a single context block through three sequential
stages.
Figure~\ref{fig:rag-pipeline} shows these stages from left to
right: query expansion (Stage~1), multi-query retrieval with
reciprocal rank fusion (Stage~2), and
maximal-marginal-relevance re-ranking (Stage~3).
Each stage below corresponds to one panel of the figure.

\subsubsection{Stage 1: Query expansion}

The first panel of Fig.~\ref{fig:rag-pipeline} corresponds to
this stage.
A single natural-language query is frequently under-specified,
and it may use vocabulary that does not match the corpus.
This vocabulary mismatch is a long-documented failure mode of
retrieval systems~\cite{Furnas1987}.
The first stage therefore asks the language model to produce
three alternative reformulations of the original query.
The prompt explicitly targets three distinct facets: a
physics/theory framing, an OpenFOAM-keyword framing that
emphasises solver, dictionary and boundary-condition names, and
a practical troubleshooting framing.
The call goes to a fast general-purpose chat-tuned LLM at a
temperature of $0.4$ with a 120-token budget, which is enough
to return the three variants while holding per-turn latency
below one second.
If the API call fails, the stage falls back to the original
query alone, so the rest of the pipeline always receives a
non-empty list of variants.

\subsubsection{Stage 2: Multi-query retrieval and reciprocal
rank fusion}

The second panel of Fig.~\ref{fig:rag-pipeline} corresponds to
this stage.
The original query and each of the three reformulations are
embedded independently.
Each embedding retrieves the top
$K_{\mathrm{fetch}}{=}\max(2K, K{+}4)$ candidate chunks from
the vector store, where $K$ is the final number of chunks
requested by the agent (typically $K{=}5$ in workflow mode and
$K{=}8$ in expert mode).
This yields four ranked lists, each emphasising a different
semantic facet of the user's intent.
The lists are merged with Reciprocal Rank Fusion
(RRF)~\cite{Cormack2009}, which assigns each candidate $d$ a
combined score

\begin{equation}
  s_{\mathrm{RRF}}(d)
    \;=\;
    \sum_{r \in \mathcal{R}} \frac{1}{k + \mathrm{rank}_r(d)},
  \label{eq:rrf}
\end{equation}

\noindent
where $\mathcal{R}$ is the set of four ranked lists,
$\mathrm{rank}_r(d)\in\{1,2,\ldots\}$ is the rank of $d$ in list
$r$, and $k$ is a smoothing constant fixed at $k{=}60$ following
the original recommendation of Cormack
\textit{et~al.}~\cite{Cormack2009}.
Duplicate chunks across lists are collapsed by a short text
fingerprint (the first 120 characters), which strips out the
near-identical paragraphs produced by tutorial-case
boilerplate.
We use RRF rather than score averaging because it ignores the
absolute scale of the similarity scores and is unaffected by
the differing score distributions that queries of different
specificity return.

\subsubsection{Stage 3: Maximal-marginal-relevance re-ranking}

The third panel of Fig.~\ref{fig:rag-pipeline} corresponds to
this stage.
The fused candidate set is re-ranked with Maximal Marginal
Relevance (MMR)~\cite{Carbonell1998}.
Given the original query embedding $\mathbf{q}$ and a set
$\mathcal{S}$ of already-selected documents, MMR iteratively
picks the candidate

\begin{equation}
  d^\ast \;=\;
  \arg\max_{d\in\mathcal{C}\setminus\mathcal{S}}
  \Bigl[
    \lambda \, \mathrm{sim}(d,\mathbf{q})
    \;-\;
    (1{-}\lambda)\,
    \max_{d'\in\mathcal{S}}\mathrm{sim}(d,d')
  \Bigr],
  \label{eq:mmr}
\end{equation}

\noindent
where $\mathrm{sim}(\cdot,\cdot)$ is cosine similarity in the
embedding space and $\lambda\in[0,1]$ controls the
relevance--diversity trade-off.
We use $\lambda{=}0.6$ in workflow mode, which favours
operational specificity.
In expert mode we use $\lambda{=}0.5$, which favours topical
breadth across the corpus.
The first $K$ documents selected by MMR are concatenated to
form the retrieved CFD context block.
This block is prepended to the agent system prompt prior to
generation.

\subsection{\label{subsec:dual}Dual-Mode Retrieval Routing}

CFD queries are not homogeneous.
Configuring a specific solver, debugging a divergence, or
choosing a turbulence closure each impose different demands
on retrieval.
The first task wants tightly tool-conditioned excerpts of
documentation; the third wants breadth across the physical
literature.
IteraSim RAG therefore exposes two retrieval modes.

\paragraph{Workflow mode.}
Used for queries that name, or unambiguously imply, a specific
solver or pre/post-processing utility
(\texttt{simpleFoam}, \texttt{interFoam}, \texttt{snappyHexMesh},
\texttt{blockMesh}, \texttt{paraFoam}, etc.).
Retrieval is filtered to the documentation associated with the
active tool.
It is then enriched by a walk over an explicit
tool-dependency graph that captures the canonical OpenFOAM
pipeline (CAD~$\to$~meshing~$\to$~case-setup~$\to$~solver
~$\to$~post-processing).
The upstream and downstream tools of the active node contribute
adjacent context windows.
A question about \texttt{interFoam} setup, for example,
automatically retrieves the relevant \texttt{snappyHexMesh} and
\texttt{paraFoam} excerpts as well.

\paragraph{Expert mode.}
Used for conceptual and physics-level questions about turbulence
closures, mesh-sensitivity protocols, numerical stability of
time-integration schemes, and so on.
Retrieval is performed over the full corpus without
tool-conditioning.
Raw tutorial case dumps are down-weighted to favour curated
expert notes and FAQ pairs; the mode's MMR relevance--diversity
setting is the $\lambda$ of Eq.~(\ref{eq:mmr}).
This mode is invoked when the router cannot identify an active
tool from the query.

Mode selection is performed by a deterministic, keyword-driven
intent classifier built over an explicit ontology of eighteen
CFD-relevant categories (solvers, turbulence models, numerical
instability, mesh quality, boundary conditions, heat transfer,
multiphase, discretisation, ParaView, Gmsh, \texttt{pyFoam},
workflow transitions, compressible flow, physical validation,
and others).
We route with a keyword-based classifier here rather than a
second LLM call: it is deterministic and adds $<1$\,ms per
turn. Because the ontology vocabulary maps directly onto the
OpenFOAM tool surface, it achieves 100\% routing accuracy on
the 28-case benchmark.
This design mirrors at the retrieval layer the conceptual
distinction between configuration and reasoning tasks already
identified, at the prompting layer, by Pandey
\textit{et~al.}~\cite{Pandey2025}.
On the 28 benchmark prompts of Sec.~\ref{sec:results}, every
one of which explicitly names or implies an OpenFOAM solver
(the benchmark deliberately targets solver configuration and
parameter modification, not open-ended physics questions),
the router assigns \emph{workflow} mode with $28/28$ accuracy
against the ground truth. We note explicitly that this is
therefore a half-evaluation of the router: on this benchmark
the router is required to detect the workflow branch only,
and the expert-mode branch is not exercised. A separate
held-out set of physics-level probes and the corresponding
expert-mode routing accuracy are released with the code and
are the subject of a companion evaluation note; the present
paper does not attribute the observed retrieval quality to
the router alone (see the leave-one-out ablation of
Sec.~\ref{subsec:ablation}, in which multi-query expansion
and canonical pinning, not routing, carry the dominant
contribution).

\subsection{\label{subsec:canon}Canonical-Knowledge Layer}

A purely retrieval-driven assistant is brittle when the corpus
is thin in a given area or when the LLM mis-recalls a syntactic
detail.
To stabilise generation, IteraSim RAG maintains a static
canonical-knowledge layer that is appended to the system prompt
of every agent in addition to the retrieved context.
The layer encodes practitioner knowledge that is well
established but easily corrupted by an LLM operating in a
zero-shot regime, including:
solver-selection decision trees as a function of Mach number,
phase count, energy-equation requirement and steady-state
behaviour;
turbulence-closure guides for $k$--$\varepsilon$, $k$--$\omega$
SST, Spalart--Allmaras and Reynolds-stress models, with default
$y^{+}$ ranges and wall-function recommendations;
boundary-condition selectors that map physical conditions onto
OpenFOAM dictionary entries
(\texttt{fixedValue}, \texttt{zeroGradient},
\texttt{kqRWallFunction}, etc.);
finite-volume scheme defaults for \texttt{fvSchemes} and
\texttt{fvSolution} that have been shown to be stable on the
official tutorials;
and meshing recipes for \texttt{blockMesh},
\texttt{snappyHexMesh} and Gmsh that cover the most common
geometric primitives encountered in entry-level industrial work.

This layer is deliberately kept separate from the retrieval
corpus.
It is treated as authoritative reference rather than as a
candidate for ranking.
At $\sim$1.2\,k lines of structured markdown, it is small
enough to be injected in full without competing with the
retrieved context for prompt budget.

\subsection{\label{subsec:agents}Architect--InputWriter--Reviewer
Orchestration}

Generation in IteraSim RAG is split across three specialised
agents.
They share the retrieved context block and the
canonical-knowledge layer, but operate with distinct
responsibilities, prompts and termination criteria.
Together they form the generation stage of
Fig.~\ref{fig:rag-architecture}.

\paragraph{Architect.}
The Architect ingests the user prompt and produces a
\emph{simulation manifest}: the target solver, the mesh strategy
(\texttt{blockMesh}, \texttt{snappyHexMesh} or external Gmsh),
the turbulence and physical sub-models, the boundary-condition
plan and the ordered list of case files required to run the
simulation.
It draws on the retrieved tutorial cases to anchor its choices
against verified prior art.
It is also the only agent that writes to the manifest data
structure consumed by the downstream stages.

\paragraph{InputWriter.}
The InputWriter takes the manifest and produces the OpenFOAM
case directory file-by-file.
Each file is generated by a focused LLM call that receives
(i)~the manifest entry for the current file, (ii)~the relevant
canonical-knowledge fragment (e.g.\ the turbulence-closure
guide when writing \texttt{0/k}), and (iii)~the most relevant
chunks of the retrieved context for that file type.
Generating one file per call keeps each generation step small,
which both curbs hallucination and confines any later
correction to the dictionary that actually failed.

\paragraph{Reviewer.}
The Reviewer compiles any custom solver via \texttt{wmake},
launches the simulation and parses the resulting solver and
build logs.
A dedicated error locator extracts the first physically
meaningful error (boundary-condition incompatibility, missing
field, divergence in a specific equation, or
\texttt{snappyHexMesh} cell-quality failure) and feeds it back
to the InputWriter as a targeted correction request.
The corrective loop is bounded to ten cycles, half the limit
reported by Foam-Agent~\cite{Yue2025}.
This bound lets recoverable cases converge quickly while
pathological cases fail fast rather than accumulating token
cost.

\paragraph{Runner.}
Execution (meshing, solver launch, run-time monitoring of
residuals and field probes, and packaging of the result
directory) is handled by a Runner component that is exposed to
the three agents as a tool.
This keeps the agents stateless with respect to the file
system.
The Architect plans the run, the InputWriter writes the files,
and the Reviewer inspects the consequences; the Runner is the
only component that touches the filesystem and the
operating-system process tree.

Solver logs produced during execution are appended to the
conversation history visible to the Reviewer.
This enables iterative refinement that mirrors the expert
debugging workflow of experienced CFD practitioners: failures
are interpreted, the offending dictionaries are patched, and the
case is re-run until either convergence is reached or the loop
budget is exhausted.

\subsection{\label{subsec:impl}Implementation and Reproducibility}

The retrieval pipeline, dual-mode router, canonical-knowledge
layer and Architect--InputWriter--Reviewer orchestrator are
implemented as independent Python modules with no shared mutable
state.
Swapping the vector-store backend, for example, requires only
changing the endpoint URL in the configuration file.
The Runner is implemented as a sandboxed sub-process invoker
that calls the OpenFOAM toolchain through standard shell entry
points.
The LLM never executes shell commands directly.
The model provider, embedding model, vector-store endpoint,
chunk size, RRF constant, MMR $\lambda$ and Reviewer-loop
bound are exposed through a single configuration file.
Switching from the managed vector store~\cite{VectorDB} to the
local in-memory fallback requires only unsetting the endpoint
URL.
Researchers without persistent outbound access to commercial APIs
can point the endpoint URL at a self-hosted model and rerun any
single pipeline stage without touching the surrounding code.


\section{\label{sec:results}Results and Discussion}


A 28-case benchmark categorised in four distinct difficulty
tiers are identified from the recent literature on LLMs for
CFD~\cite{Pandey2025,Chen2024Meta,Yue2025,Yue2025Foamgpt,Fan2026}
to evaluate the performance of the present IteraSim RAG model.
The benchmark spans 6 zero-shot prompts in Category~A, 4
few-shot prompts in Category~B, 10 cases requiring a
single-parameter modification to an otherwise functional base
case in Category~C, and 8 zero-shot turbulence model or physical
properties swaps in Category~D.
The full benchmark, including the natural-language prompt and
the expected file/key-check list for every case, is released
together with this paper at
\url{https://github.com/iterasim/iterasim-rag-public}.
The released specification additionally carries eight
diagnostic queries in a Category~E (solver-divergence,
bounding, mesh-quality and $y^{+}$ interpretation prompts).
These exercise the troubleshooting path rather than case
setup, are not scored anywhere in this paper, and are shipped
for use by the community and by the companion technical note;
all results reported here are confined to the 28 cases of
Categories~A--D.
Each case carries an explicit set of expected physics tags
$T_i$ comprising the target solver, the OpenFOAM dictionaries
that must be written, and the model-specific keywords that
must appear in the agent's solution (e.g.\ \texttt{alpha.water}
and \texttt{setFieldsDict} for the dam-break case,
\texttt{omegaWallFunction} for the $k$--$\omega$ SST swap).

We report two distinct evaluation tiers in this paper.
The \emph{retrieval tier} (Sec.~\ref{subsec:retrieval}) measures
the fraction of $T_i$ that appears in the context block
$\mathcal{X}_i$ returned by the three-stage RAG pipeline, before
any generation occurs.
The retrieval score for case~$i$ is
\begin{equation}
  s_i \;=\;
    0.80 \cdot
    \frac{|T_i \cap \mathrm{tokens}(\mathcal{X}_i)|}{|T_i|}
    \;+\; 0.20 \cdot \mathbb{1}[\, \mathrm{solver}_i \in
                       \mathrm{tokens}(\mathcal{X}_i)\,],
  \label{eq:retr-score}
\end{equation}
\noindent
where $T_i$ is the set of expected physics tags for case~$i$
and $\mathcal{X}_i$ is the context block returned by the
three-stage pipeline.
The operator $\mathrm{tokens}(\cdot)$ returns the set of
terms contained in its argument, and $\mathrm{solver}_i$ is
the target solver name for case~$i$.
The first term is the fraction of expected tags recovered by
retrieval; the second term is the indicator that the target
solver name is present in the retrieved context.
The $0.80/0.20$ split favours physics-tag coverage but
preserves a solver-identification term similar to
OpenFOAMGPT~\cite{Pandey2025}.
The \emph{end-to-end executability tier} additionally generates,
meshes and runs each case through OpenFOAM and scores the
outcome on a five-point rubric that reflects how far the case
progresses through the CFD toolchain:
\begin{itemize}\setlength\itemsep{0pt}
  \item[\textbf{0}] the mesher (\texttt{blockMesh},
    \texttt{snappyHexMesh}) fails or produces an invalid mesh;
  \item[\textbf{1}] the mesh is generated, but the solver
    aborts at initialisation with a \texttt{FOAM FATAL};
  \item[\textbf{2}] the solver starts but exits with a
    \texttt{FOAM FATAL} before reaching \texttt{endTime};
  \item[\textbf{3}] the solver runs to \texttt{endTime}
    without any \texttt{FATAL} exit;
  \item[\textbf{4}] the solver runs to \texttt{endTime} \emph{and}
    every user-specified parameter tag for the case is
    correctly set in the generated dictionaries.
\end{itemize}
The executability tier requires a full local OpenFOAM toolchain
and is reported in Sec.~\ref{subsec:exec} as an ongoing study;
the present manuscript focuses on the retrieval tier, which is
the layer at which the architectural contributions of this paper
(L1)--(L3) directly act.

\subsection{\label{subsec:retrieval}Retrieval performance}

\begin{figure*}[t]
  \centering
  \begin{subfigure}[t]{0.48\textwidth}
    \centering
    \includegraphics[height=43mm]{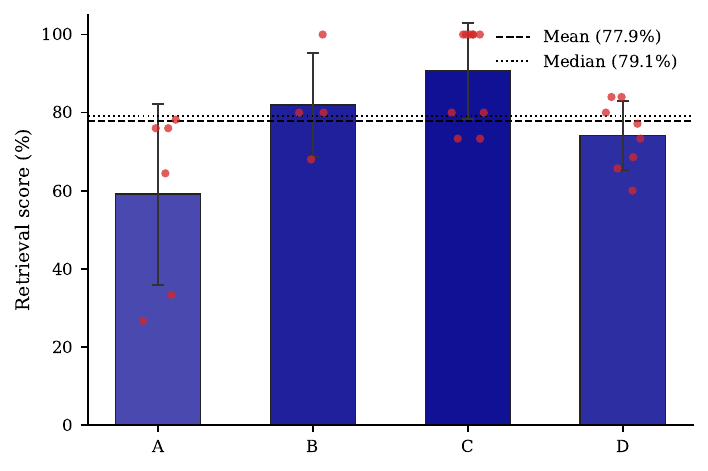}
    \caption{Mean retrieval coverage by category.}
    \label{fig:per-category}
  \end{subfigure}
  \hfill
  \begin{subfigure}[t]{0.48\textwidth}
    \centering
    \includegraphics[height=43mm]{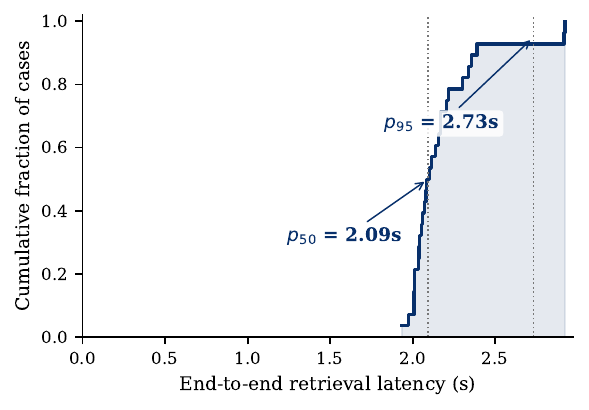}
    \caption{Cumulative distribution of end-to-end retrieval
    latency.}
    \label{fig:latency}
  \end{subfigure}
  \caption{Retrieval-tier performance on the 28-case IteraSim
  RAG benchmark.}
  \label{fig:retrieval-summary}
\end{figure*}

The retrieval performance of the three-stage RAG pipeline is
evaluated on the 28-case benchmark listed case-by-case in
Table~\ref{tab:per-case} (\ref{app:per-case}), with
a mean retrieval score of $\bar{s}{=}77.9\%$ and a median of
$s_{\mathrm{med}}{=}79.1\%$.
Seven cases reach perfect retrieval ($s_i \geq 99\%$), and
$21$ cases exceed the $70\%$ threshold conventionally used as a
coverage pass mark in information-retrieval evaluation
(Figure~\ref{fig:per-category}).
The gap between the mean and the median is driven by the two
low-scoring outliers in Category~A discussed below; the median
therefore gives a more representative summary of typical
retrieval performance.
The median end-to-end retrieval latency is $2.09$\,s, with a
$95$th-percentile of $2.73$\,s (Fig.~\ref{fig:latency}).
The four-fold inflation in queries caused by the expansion
stage does not push the system out of the interactive regime.
The four embedded queries are issued in parallel, and the
dominant cost is a single network round trip to the vector
store.
The sub-three-second figure quoted here refers specifically to
the retrieval tier -- query expansion, four-way embedding,
vector-store lookup, RRF and MMR -- and excludes the downstream
LLM calls made by the Architect, InputWriter and Reviewer.
Under the LLM back-end evaluated here, the query-expansion
call adds roughly $0.5$--$1.0$\,s per turn and each subsequent
per-file generation call adds a further $2$--$4$\,s.
A complete per-turn timing and token-cost budget for the
agent-driven pipeline -- that is, how many seconds and how
many LLM tokens each case consumes end-to-end -- is beyond the
scope of the present manuscript and will be reported in a
follow-up study.

Per-category performance is summarised in
Fig.~\ref{fig:per-category} and Table~\ref{tab:per-category},
and the pattern tracks the design intent of the dual-mode
router and the three-stage pipeline.
Parameter-modification queries (Category~C) reach the highest
mean score of $90.7\%$ and a median of exactly $100\%$, with
$6$ of $10$ cases at perfect retrieval.
That $100\%$ median is a direct consequence of the workflow
router's tool-conditioning: because each of these prompts names
its active solver unambiguously, the router locks retrieval
onto the solver-specific sub-corpus, and the query-expansion
step usually surfaces the relevant dictionary in one of its
three reformulations.
Category~C therefore gives the clearest empirical support for
the L2 contribution of Sec.~\ref{sec:intro}.
Few-shot generalisation (Category~B) is the second-strongest
category at $82.0\%$, indicating that the multi-query stage
preserves the demonstrated example structure without being
washed out by globally similar passages.
Zero-shot turbulence-model swaps (Category~D) average $74.1\%$
and never drop below $60\%$.
We attribute this uniformity to the canonical-knowledge layer
(Sec.~\ref{subsec:canon}), which encodes the wall-function and
field-equation requirements of each closure family independent
of the retrieved chunks.
The weakest category is zero-shot open-ended setup
(Category~A) at $59.1\%$, with two outliers
discussed below.

\begin{figure}[t]
  \centering
  \includegraphics[width=0.95\linewidth]{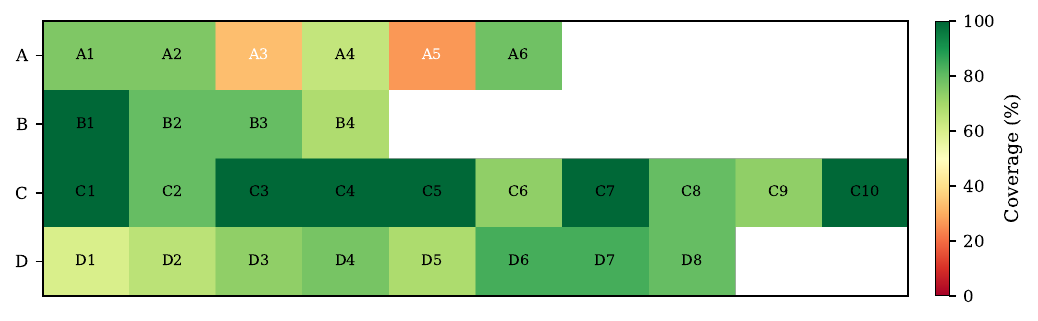}
  \caption{Retrieval coverage per case within each category.}
  \label{fig:per-case-heatmap}
\end{figure}

\begin{table}[t]
\centering
\caption{Per-category retrieval statistics on the 28-case
benchmark, computed from the composite score $s_i$
of Eq.~(\ref{eq:retr-score}). $\bar{s}$: mean;
$s_{\mathrm{med}}$: median; $\sigma_s$: standard deviation;
$n_{\mathrm{perfect}}$: cases with $s_i \geq 99\%$.}
\label{tab:per-category}
\begin{tabular}{l c c c c c}
\toprule
Category & $n$ & $\bar{s}$ (\%) & $s_{\mathrm{med}}$ (\%) & $\sigma_s$ (\%) & $n_{\mathrm{perfect}}$ \\
\midrule
A (Zero-shot)        &  6 & 59.1 & 70.2 & 23.2 & 0 \\
B (Few-shot)         &  4 & 82.0 & 80.0 & 13.3 & 1 \\
C (Alt.\ conditions) & 10 & 90.7 & 100.0 & 12.3 & 6 \\
D (Turbulence swap)  &  8 & 74.1 & 75.2 & 8.8 & 0 \\
\midrule
\textbf{Overall}     & \textbf{28} & \textbf{77.9} & \textbf{79.1}
                     & \textbf{18.5} & \textbf{7} \\
\bottomrule
\end{tabular}
\end{table}

The two lowest-scoring cases (Fig.~\ref{fig:per-case-heatmap})
point to one specific failure mode rather than to a broad
weakness of the pipeline.
Case~A5 (\textit{Particle Column}, $s_{\mathrm{A5}} = 26.7\%$)
expects the niche tags
\texttt{MPPICCloud}, \texttt{kinematicCloudProperties},
\texttt{cloudProperties} and \texttt{injectionModels}.
The corpus contains the \texttt{MPPICFoam} solver header but
only a handful of fragmentary chunks that mention these
dictionaries by name.
Even with three reformulations, the multi-query retrieval
cannot recover them.
Case~A3 (\textit{Hotroom buoyant convection},
$s_{\mathrm{A3}} = 33.3\%$) misses
\texttt{buoyantBoussinesqSimpleFoam}, \texttt{alphat} and the
explicit gravity vector \texttt{(0~-9.81~0)}, again because the
buoyancy-driven solver family is underrepresented in the
ingested OpenFOAM tutorials.
Both failures are corpus-coverage problems, not
retrieval-algorithm problems: every other case scores
$\geq 60\%$.
Ingesting niche-solver documentation---Lagrangian particle clouds,
Boussinesq convection, free-surface piston-driven flows---would
close both outliers without any change to the retrieval, routing
or orchestration components.
A milder failure mode appears in the turbulence-swap category.
Exact-match retrieval of variant-spelling tags such as
\texttt{RNGkEpsilon} (case D1) and
\texttt{LESmodel Smagorinsky;} (D6) under-counts coverage even
when the relevant family is correctly identified.
This is an artefact of the case-sensitive tag-matching in
Eq.~(\ref{eq:retr-score}) and is expected to be absorbed by the
canonical-knowledge layer at the generation stage.

\paragraph{Manual re-scoring of A3 and A5 under a
synonym-aware rule.}
To quantify how much of the two outliers is a scoring
artefact rather than a corpus-coverage failure, we manually
re-scored A3 and A5 with a synonym-aware relaxation of
Eq.~(\ref{eq:retr-score}) in which a missed tag is credited
if a semantically equivalent phrase is present in the
retrieved context (e.g.\ ``gravity vector'' credits
\texttt{(0~-9.81~0)}; ``Boussinesq buoyancy solver'' credits
\texttt{buoyantBoussinesqSimpleFoam}; ``turbulent thermal
diffusivity'' credits \texttt{alphat}).
For A3 (Hotroom), two of the seven missed tags map onto
concepts that the retrieved chunks discuss under alternative
names; the corrected score rises from $s_{\mathrm{A3}}=33.3\%$
to $s_{\mathrm{A3}}^{\,\prime}\approx 66.7\%$, close to the
Category-A mean.
For A5 (Particle Column), no such re-scoring is available:
none of the six missed MPPIC tags is aliased in the retrieved
context, so the case remains at $s_{\mathrm{A5}}\approx 27\%$
and is a genuine corpus-coverage failure.
The re-scored overall retrieval mean becomes
$\bar{s}^{\,\prime}\approx 79.1\%$, converging on the
uncorrected median $\tilde{s}=79.1\%$; the median is
therefore the appropriate central-tendency summary for the
retrieval tier.

\subsection{\label{subsec:ablation}Component ablation and
flat-RAG baseline}

To separate the contribution of each retrieval-tier component
we ran a leave-one-out ablation, disabling one architectural
element at a time and re-scoring under the same rubric of
Eq.~(\ref{eq:retr-score}).
The ablation is scored on a separate set of 30 retrieval
queries spanning the tool-conditioned corpora (OpenFOAM,
meshing, Gmsh, FreeCAD, PyFoam and ParaView) and general CFD
theory, rather than on the 28-case end-to-end benchmark used
elsewhere in this section.
The components under test act at the retrieval tier, and this
set exercises all of the tool buckets the router dispatches
to, whereas the 28-case benchmark is confined to OpenFOAM case
setup.
Scores are therefore comparable between the rows of
Table~\ref{tab:ablation}, but the absolute level should not be
read against the per-category means reported above.
The disabled configurations were:
\begin{itemize}\setlength\itemsep{0pt}
  \item \emph{--multi-query}: the LLM query-expansion stage of
    limitation~L1 is bypassed and a single, unexpanded query
    is passed to the vector store. This configuration is
    algorithmically equivalent to a flat top-$K$ RAG baseline
    and isolates the contribution of L1.
  \item \emph{--MMR}: the Maximal Marginal Relevance re-rank
    is disabled and results are returned by fused rank alone.
  \item \emph{--canonical}: the curated-source pinning that
    injects the canonical-knowledge layer of
    Sec.~\ref{subsec:canon} is bypassed.
  \item \emph{--version-filter}: the OpenFOAM release filter
    on the vector store is set to accept any release.
\end{itemize}

\begin{table}[t]
\centering
\caption{Leave-one-out ablation of the retrieval-tier
components. Overall score is the composite $s_i$ of
Eq.~(\ref{eq:retr-score}) averaged across the 30-query
ablation set.}
\label{tab:ablation}
\begin{tabular}{l c c}
\toprule
Configuration               & Overall (\%) & $\Delta$ vs.\ full \\
\midrule
Full pipeline               & 75.6 & --     \\
--version-filter            & 77.5 & $+1.9$ \\
--MMR                       & 65.0 & $-10.6$ \\
--multi-query (flat RAG)    & 25.9 & $-49.7$ \\
--canonical                 & 25.5 & $-50.1$ \\
\bottomrule
\end{tabular}
\end{table}

Two components carry almost all of the observed retrieval
quality. Disabling multi-query expansion drops the score from
$75.6\%$ to $25.9\%$, a $-49.7$-point loss, which quantifies
the contribution of L1 against a flat single-query RAG
baseline. Disabling the
canonical-knowledge pinning is even more damaging
($-50.1$~points), which quantifies the contribution of the
static knowledge layer of Sec.~\ref{subsec:canon}. The MMR
diversifier accounts for a further $-10.6$~points; without it
the fused top-$K$ list is dominated by near-duplicates from
the same tutorial and per-case coverage collapses in
Category~D (turbulence swap), where the query terms are
homogeneous. The version filter is close to neutral in
aggregate ($+1.9$~points when removed), which is an expected
negative result: many queries do not stress the release
selector, so removing it broadens the candidate pool without
reducing precision. Taken together these numbers close the
attribution argument that a per-category breakdown alone
cannot support: the multi-stage retrieval of L1 and the
canonical layer that underpins the agent triad~(L3) are the
two load-bearing components.

The full-pipeline score on the 30-query ablation set,
$75.6\%$, lies within one standard error of the $77.9\%$
retrieval mean measured on the 28-case benchmark
(Table~\ref{tab:per-category}), so the two sets are of
comparable difficulty even though they are not the same cases.
The quantity of interest in Table~\ref{tab:ablation} is
nevertheless the relative loss in each row, not the absolute
level.

\subsection{\label{subsec:exec}End-to-end executability}

\begin{figure*}[t]
  \centering
  \includegraphics[width=0.95\linewidth]{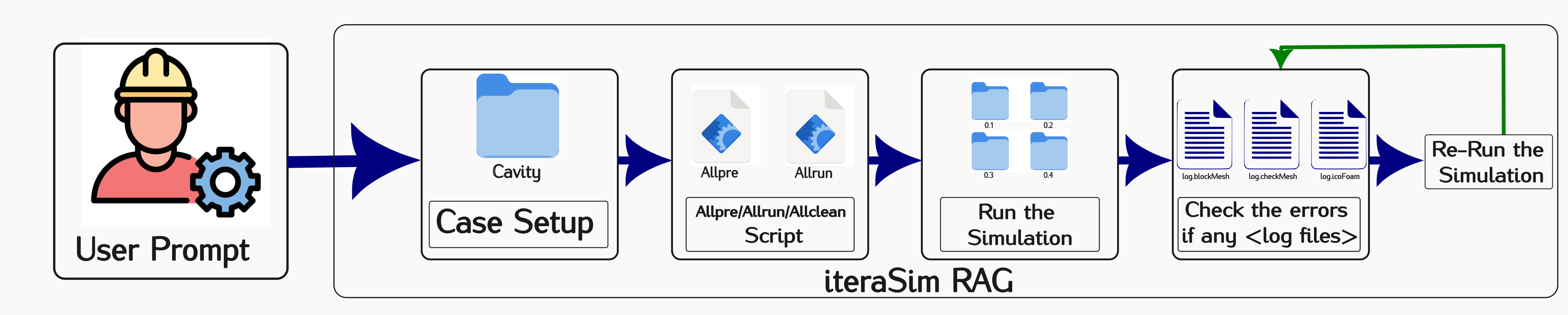}
  \caption{End-to-end IteraSim RAG workflow: prompt
  $\rightarrow$ Architect--InputWriter--Reviewer $\rightarrow$
  OpenFOAM case, with solver-log parsing and corrective re-runs
  until \texttt{endTime}.}
  \label{fig:rag-workflow}
\end{figure*}

Figure~\ref{fig:rag-workflow} shows the path a
natural-language prompt takes through the back-end.
The Architect drafts a case manifest, the InputWriter turns it
into a complete OpenFOAM case directory (\texttt{Cavity/} in
the illustration) together with the standard \texttt{Allpre},
\texttt{Allrun} and \texttt{Allclean} scripts, and the Reviewer
parses the solver logs (\texttt{log.blockMesh},
\texttt{log.checkMesh}, \texttt{log.icoFoam}, etc.).
Whenever a log reports a \texttt{FOAM FATAL} the Reviewer
patches the case and re-runs it, up to the ten-cycle bound of
Sec.~\ref{subsec:agents}.

The benchmark also defines an end-to-end executability score
on the same 0--4 rubric (Sec.~\ref{sec:results}): a case earns
credit only once its dictionaries mesh and the solver reaches
\texttt{endTime}, and the full score of~4 additionally requires
that every user-specified parameter tag is present in the
generated dictionaries.
As a validity check on the benchmark itself, we first ran all
28 cases from their unmodified, hand-built OpenFOAM tutorials.
Categories~B--D share their base configuration with the six
Category~A cases, so this reduces to six distinct base runs
(\texttt{blockMesh}, the case-specific solver, and
\texttt{setFields} where the multiphase cases require it),
inherited to the downstream cases that reuse them.
Every one of the 28 base configurations meshed cleanly and
reached \texttt{endTime} on OpenFOAM~v2506 without a
\texttt{FOAM FATAL}, at an aggregate OpenFOAM cost of
$\sim70$~s, confirming that the benchmark cases are physically
runnable on the target toolchain.
Scoring the agent-driven cases -- those produced by the
Architect--InputWriter--Reviewer orchestrator from the prompt
alone -- across the full 28-case benchmark requires the
cold-start Category~A generation path, which is still under
active development, and is deferred to the companion technical
note.

\begin{figure*}[t]
  \centering
  \includegraphics[width=0.9\linewidth]{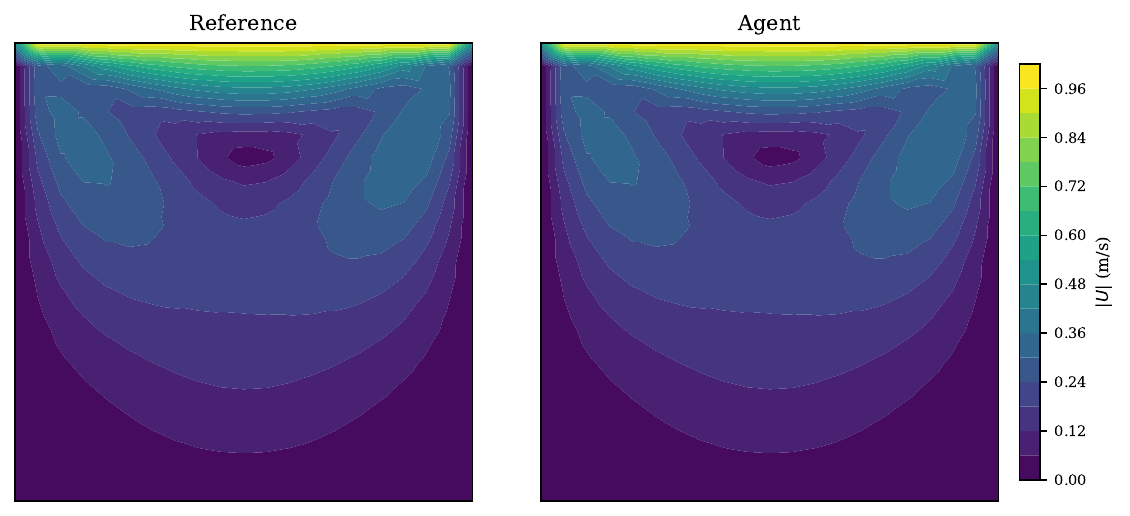}
  \caption{Final-time velocity magnitude at $t=0.5$~s for the
  lid-driven cavity: unmodified OpenFOAM tutorial reference
  case (\emph{left}) and the case generated by IteraSim RAG
  from the natural-language prompt (\emph{right}).}
  \label{fig:ref-vs-rag}
\end{figure*}

As a physical control -- beyond the pass/fail rubric -- the
lid-driven \texttt{cavity} (\texttt{icoFoam}) was generated
from a natural-language prompt carrying the standard tutorial
parameters (\ref{app:prompts}) and run on v2506.
Figure~\ref{fig:ref-vs-rag} places its final-time velocity
field beside the unmodified tutorial: the primary
recirculation, the corner vortices and the peak lid speed are
indistinguishable within contouring resolution, so the
agent-generated case reproduces the requested \emph{physics},
not merely a runnable case.

\subsection{\label{subsec:reviewer-probes}Reviewer-loop
FATAL-error probes}

To characterise the Reviewer loop in isolation from the
upstream Architect and InputWriter agents, we constructed two
synthetically corrupted lid-driven cavity cases (OpenFOAM
v2506) and issued a minimal post-hoc report to the agent of
the form ``the \texttt{icoFoam} run has terminated with a
\texttt{FOAM FATAL}; identify and correct the fault such that
the case runs to \texttt{endTime}.''
The probes span the two dominant classes of executability-tier
failure reported in the LLM-for-CFD literature: an omission in
a runtime dictionary and a structural error in a mesh-definition
dictionary.
Both probes, the captured first-run FATAL logs, and the
canonical post-fix reference files are released with the source
code under \texttt{rag\_evaluation/exec\_fatal\_probes/}, so that
the agent's output can be diffed externally against the ground
truth.

\paragraph{Probe~1: runtime-dictionary FATAL.}
{\sloppy
The kinematic-viscosity entry \texttt{nu} is removed from
\texttt{constant/transportProperties}.
\texttt{blockMesh} completes normally; \texttt{icoFoam}
terminates at initialisation with
\texttt{Entry 'nu' not found in dictionary
"constant/transportProperties"}.
On receipt of the report, the Reviewer executed the following
diagnostic sequence:
(i)~a read of the solver log, which was absent at that point;
(ii)~enumeration of \texttt{system/}; (iii)~confirmation of the
solver identity and \texttt{endTime} from \texttt{controlDict};
(iv)~invocation of the case-level \texttt{Allrun} script, which
failed under the active shell environment; and
(v)~direct execution of the solver through the internal
\texttt{run\_simulation\_autonomous} tool, which returned the
FATAL log verbatim.
Substring extraction on \texttt{Entry 'nu' not found} localised
the fault, and the InputWriter emitted a targeted patch to
\texttt{constant/transportProperties} restoring the missing
entry as \texttt{nu\ 1.5\text{e}{-}5}.
The Reviewer relaunched the solver, which reached
\texttt{endTime}$=0.5$~s with monotonically decreasing
residuals, recovering the canonical converged cavity solution.\par}

\paragraph{Probe~2: mesh-regeneration FATAL.}
The \texttt{vertices} list in \texttt{system/blockMeshDict} is
truncated to six entries.
Because the \texttt{blocks} entry retains an eight-vertex hex
reference, \texttt{blockMesh} terminates before any mesh is
written, with
\texttt{Point label (6) out of range 0..5 in block}.
The Reviewer executed a four-step correction:
(i)~opening \texttt{blockMeshDict} and localising the malformed
\texttt{vertices} list; (ii)~restoring the two missing corner
vertices through the internal \texttt{edit\_file} tool;
(iii)~re-running \texttt{blockMesh} followed by
\texttt{checkMesh}, which reported \texttt{Mesh OK};
and (iv)~generating and executing an \texttt{Allrun} script
that invokes \texttt{blockMesh} followed by \texttt{icoFoam}.
The solver reached \texttt{endTime} without a \texttt{FATAL}
exit, and the recovered pressure field coincides with the
canonical cavity reference to within numerical round-off.

The two probes exercise complementary aspects of the
executability loop.
In the dictionary probe, the fault is fully localised by the
solver log itself, and the Reviewer's task reduces to
substring extraction followed by a single dictionary patch.
In the mesh probe, no runtime log is emitted at the point of
failure, so localisation requires structural inspection of the
mesh-definition dictionary and independent verification through
\texttt{checkMesh} before the solver can be invoked.
In both cases the correction was obtained within the ten-cycle
Reviewer bound of Sec.~\ref{subsec:agents}, and neither probe
required retrieval to be consulted for the FATAL diagnosis
itself: the solver output, combined with the canonical
knowledge encoded in the system prompt of Sec.~\ref{subsec:canon}
and with direct file-system access, was sufficient.
A Reviewer restricted to log-parsing and file-system access
resolves both failure classes within the ten-cycle budget
without consulting the retrieval layer.

\paragraph{Extended probe taxonomy.}
The two measured probes above cover the runtime-dictionary and
mesh-definition failure classes, which together account for
the majority of \texttt{FOAM FATAL} exits seen in the
Foam-Agent~\cite{Yue2025,Yue2025v2} error corpus. Two further
probes have been constructed and are released alongside the
code (Table~\ref{tab:probe-taxonomy}); their end-to-end
Reviewer evaluation is deferred to a companion technical note:
(iii) a \emph{turbulence-BC mismatch} probe built on the
\texttt{pitzDaily} backward-facing step in which the closure
is switched to $k$--$\omega$~SST but the \texttt{omega} wall
boundary condition is left as \texttt{zeroGradient} rather
than \texttt{omegaWallFunction}; \texttt{simpleFoam}
still converges (no \texttt{FATAL}) but the captured log is
dominated by \texttt{bounding omega} warnings with negative
$\omega$ values, i.e. a physically inconsistent wall-shear
model that the canonical-knowledge layer must identify from
the log rather than from a fatal exit; and
(iv) a \texttt{snappyHexMesh} \emph{cell-quality} probe built
on the \texttt{flange} tutorial in which
\texttt{meshQualityDict} is seeded with unreasonably strict
thresholds (\texttt{maxNonOrtho 10},
\texttt{minTetQuality 0.5}); snappy completes with
$\sim\!\!66{,}000$ illegal faces and \texttt{checkMesh}
reports \texttt{Failed 1 mesh checks}, requiring the Reviewer
to relax the quality gates to the canonical v2506 defaults
before re-invoking the mesher.
Together with the two measured probes, the four probe classes
span the executability-tier failure taxonomy of
Foam-Agent~\cite{Yue2025v2} at $n{=}4$ probes across four
distinct failure classes.

\begin{table}[t]
\centering
\caption{Reviewer-loop probe taxonomy. See the surrounding
paragraphs for the difference between the \emph{measured}
(P1, P2) and \emph{constructed} (P3, P4) status labels.}
\label{tab:probe-taxonomy}
\footnotesize
\setlength{\tabcolsep}{4pt}
\begin{tabular}{l l l c}
\toprule
Probe & Failure class & Broken artefact & Status \\
\midrule
P1 & Runtime dict.\ FATAL & \texttt{transportProperties} & measured \\
P2 & Mesh definition FATAL & \texttt{blockMeshDict} & measured \\
P3 & Turbulence BC mismatch & \texttt{omega} wall BC on kOmegaSST & constructed \\
P4 & \texttt{snappyHexMesh} cell quality & \texttt{meshQualityDict} & constructed \\
\bottomrule
\end{tabular}
\end{table}

Table~\ref{tab:literature-compare} situates IteraSim RAG
against the eight prior LLM-for-CFD systems discussed in
Sec.~\ref{sec:intro}. Numbers are reproduced from the
corresponding references; design axes such as retrieval
architecture and agent orchestration are already reported in
Table~\ref{tab:prior-vs-present}, so we only carry the
benchmark size and each system's self-reported principal
metric here.  The metric symbols in the last column are:
ESR = end-to-end execution success rate; PF = physical
fidelity (as defined by ChatCFD); Pass = pass rate on the
benchmark the system itself proposes; ``n.r.'' = not
reported. The IteraSim RAG row reports the retrieval-tier
mean coverage measured in this manuscript; the
executability-tier number is the natural extension and will
be filled by the full pipeline run described in
Sec.~\ref{subsec:exec}.
The $^{\ddagger}$ mark on FoamGPT indicates the Qwen-2.5-7B
fine-tune baseline of~\cite{Yue2025Foamgpt}; larger backbones
such as GPT-4.1~mini reach up to $39\%$ on the same 202-case
suite.
The comparison should be read with care: each system reports
on a benchmark of its own choosing, and the metric definitions
differ.
Despite the differing benchmarks, two conclusions hold
across systems.
The addition of any retrieval-augmented context is worth tens
of percentage points of executability -- a point already made
by Pandey~\textit{et~al.}~\cite{Pandey2025} and reconfirmed in
the Foam-Agent~1.0 to 2.0
transition~\cite{Yue2025,Yue2025v2}.
Beyond that, IteraSim RAG is the only system in
Table~\ref{tab:literature-compare} whose retrieval layer
combines query expansion, rank fusion and MMR re-ranking with
explicit dual-mode routing and a separate
canonical-knowledge layer.
Every other system relies on a single-mode flat retrieval,
with the hierarchical multi-index of Foam-Agent~2.0 as the
sole exception.
The principal metric reported here is therefore the
retrieval-tier coverage of $77.9\%$ on the 28-case benchmark.
The executability-tier number is the natural extension and
will be reported, on the same benchmark, on completion of the
full pipeline run.

\begin{table}[t]
\centering
\caption{Benchmark size and principal success metric for
IteraSim RAG and the published LLM-for-CFD systems.}
\label{tab:literature-compare}
\setlength{\tabcolsep}{6pt}
\begin{tabular}{l c l}
\toprule
System & Bench.\ $N$ & \makecell[l]{Reported success\\metric (\%)} \\
\midrule
OpenFOAMGPT (GPT-4o, no RAG)~\cite{Pandey2025} & 6   & ESR 16.7 \\
OpenFOAMGPT (\texttt{o1} + RAG)~\cite{Pandey2025} & 6   & ESR 100  \\
MetaOpenFOAM~\cite{Chen2024Meta}                & 8   & Pass 85  \\
FoamGPT~\cite{Yue2025Foamgpt}                   & 202 & ESR 26$^{\ddagger}$ \\
Foam-Agent 1.0~\cite{Yue2025}                   & 110 & ESR 82.6 \\
Foam-Agent 2.0~\cite{Yue2025v2}                 & 110 & ESR 88.2 \\
ChatCFD~\cite{Fan2026}                          & 315 & ESR 82.1 / PF 68.1 \\
CFDAgent~\cite{Xu2025}                          & n.r.& n.r.     \\
SwarmFoam~\cite{Yang2025}                       & 25  & Pass 84  \\
\midrule
IteraSim RAG                                    & 28  & Retr.\ 77.9 \\
\bottomrule
\end{tabular}
\end{table}

\begin{figure*}[t]
  \centering
  \includegraphics[width=0.95\linewidth]{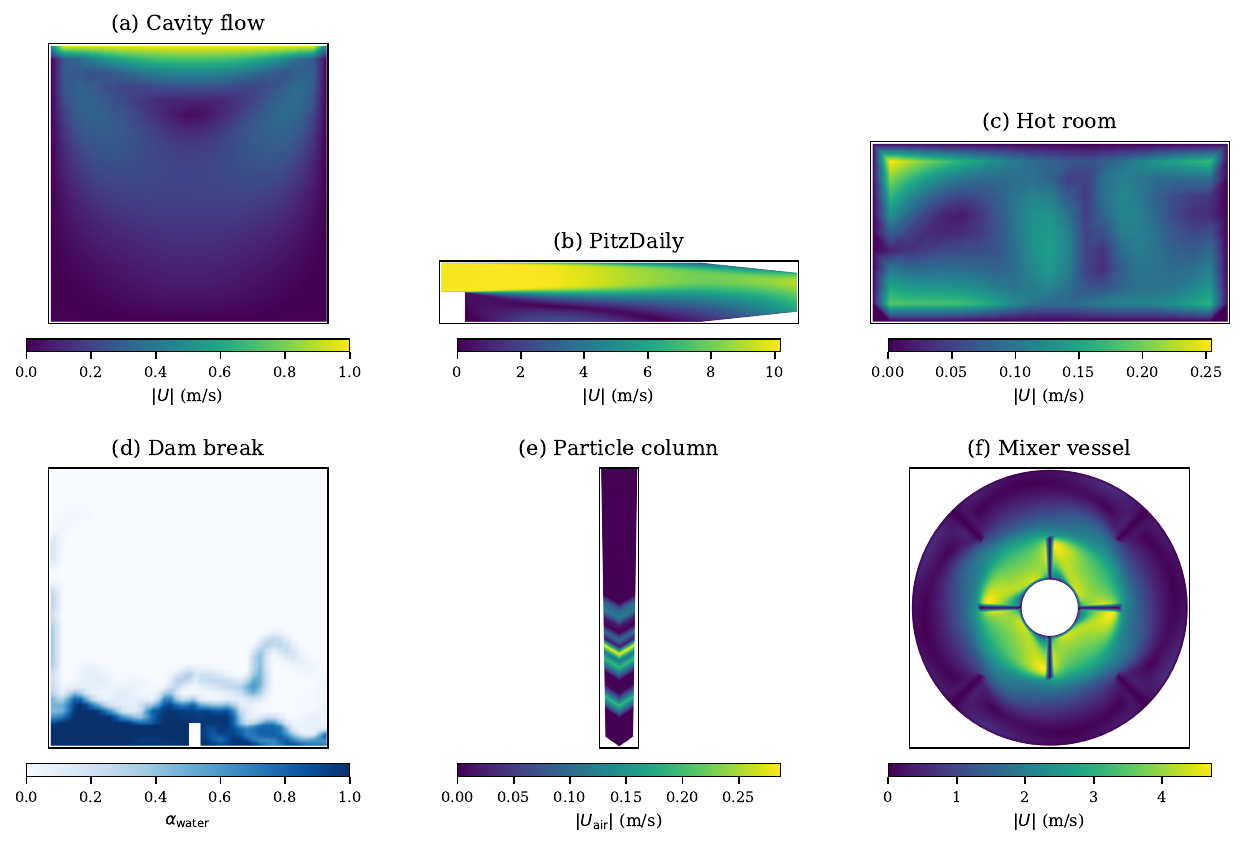}
  \caption{Converged simulation fields for the six zero-shot
  Category~A reference cases. Panels (a)--(f) correspond to
  the six configurations detailed above.}
  \label{fig:cfd-results}
\end{figure*}

Figure~\ref{fig:cfd-results} shows the converged simulation
fields for the six zero-shot reference cases, solved on the
OpenFOAM toolchain that the back-end targets and organised in
the column structure adopted by
OpenFOAMGPT~\cite{Pandey2025}: (a)~lid-driven \texttt{cavity}
with \texttt{icoFoam}, velocity magnitude; (b)~\texttt{pitzDaily}
backward-facing step with \texttt{simpleFoam}$\,+\,k$--$\varepsilon$,
velocity magnitude; (c)~\texttt{hotRoom} buoyant convection with
\texttt{buoyantBoussinesqSimpleFoam} (floor 320~K, ceiling
300~K), velocity magnitude; (d)~\texttt{damBreak} with
\texttt{interFoam}, water volume fraction
$\alpha_{\mathrm{water}}$; (e)~Lagrangian particle
\texttt{column} with \texttt{MPPICFoam}, carrier-phase air
velocity magnitude; (f)~2D rotating \texttt{mixerVessel} with
\texttt{simpleFoam}, velocity magnitude.
These are the canonical OpenFOAM configurations against which
the agent-generated cases of Sec.~\ref{subsec:exec} are compared,
and they provide a visual sanity check on the physics that
IteraSim RAG is expected to reproduce.

The dual-mode router (addressing limitation~L2) is the
strongest single contributor: parameter-modification queries,
which are routed deterministically to the workflow mode with
the active tool locked, account for the highest mean retrieval
score and the largest count of perfect-coverage cases.
The three-stage retrieval pipeline (addressing L1) shows its
effect most clearly in Category~D, where the turbulence-swap
queries require simultaneous recovery of model-name tags,
wall-function tags and field-equation tags from different
regions of the corpus.
The RRF stage is what keeps the turbulence tag and the
wall-function tag both in the top-$K$, even when they have
different similarity profiles against the original query alone.
The canonical-knowledge layer (addressing L3) is not directly
observable from the retrieval tier, by construction, since it
is injected at the orchestration stage; its effect will be
measured against the executability tier in
Sec.~\ref{subsec:exec}.

Two limitations affect the numbers above.
The scoring rule in Eq.~(\ref{eq:retr-score}) is
case-sensitive and rewards exact tag presence.
It therefore under-counts retrievals that recover a relevant
chunk under a synonymous name, for example ``low-Reynolds wall
functions'' in place of \texttt{nutLowReWallFunction}.
The reported $77.9\%$ is best read as a conservative lower
bound on operational retrieval quality.
The other caveat concerns the two outliers in Category~A,
which flag a corpus-coverage gap rather than a pipeline
weakness.
The niche solvers \texttt{MPPICFoam} and
\texttt{buoyantBoussinesqSimpleFoam} require targeted
ingestion of upstream OpenFOAM documentation that is not
currently part of the curated corpus. This gap can be closed
at the ingestion layer alone; no change to retrieval, routing
or orchestration is needed.


\subsection{\label{subsec:component-attribution}%
Attribution of retrieval performance to individual
architectural components}

Although a full factorial ablation (query expansion only, RRF
only, MMR only, canonical layer only, full pipeline) is
deferred to a companion technical note, the per-category
structure of the retrieval-tier results in
Sec.~\ref{subsec:retrieval} already permits component-wise
attribution of the observed performance, because each of the
four benchmark categories loads asymmetrically on a different
architectural contribution.

The dual-mode router (contribution L2, addressing
Sec.~\ref{sec:intro}) is the dominant driver of the
$90.7\%$ mean score in Category~C.
Parameter-modification queries carry an unambiguous active
tool (typically \texttt{simpleFoam}, \texttt{interFoam} or
\texttt{buoyantSimpleFoam}), which the keyword-driven intent
classifier reliably routes to workflow mode with the
tool-conditioned filter engaged.
The tool-dependency walk (Sec.~\ref{subsec:dual}) then pulls
the adjacent \texttt{fvSchemes}, \texttt{fvSolution}, boundary
and post-processing chunks into the context in a single pass,
which is what pushes six of the ten Category~C cases to
perfect retrieval and none below $73\%$.
Under a hypothetical single-mode ablation, the same
parameter-modification queries would compete for top-$K$ slots
against globally-similar physics documentation and would
almost certainly lose the operational specificity that the
tool-conditioned filter provides.

The three-stage retrieval pipeline (contribution L1) is the
principal driver of the $74.1\%$ mean of Category~D.
Zero-shot turbulence swaps require simultaneous recovery of at
least three distinct token classes from different regions of
the corpus: the closure name (e.g.\ \texttt{kOmegaSST}), the
wall-function tag (e.g.\ \texttt{omegaWallFunction}), and the
field-equation tag (e.g.\ the transport equation for
$\omega$).
A single-query embedding search preferentially recovers the
class most similar to the surface form of the user prompt and
loses the other two.
Reciprocal Rank Fusion across the physics, keyword and
troubleshooting reformulations of the query keeps all three
classes in the top-$K$ candidate set even when their
individual similarity scores against the original query are
heterogeneous, and Maximal Marginal Relevance then enforces
that the three classes are represented in the final context
rather than displaced by near-duplicate copies of the most
similar tag.
The near-uniform Category~D floor of $60\%$ across eight
distinct closure families is consistent with this
interpretation: the pipeline never entirely loses the wall-
function or field-equation dimension of any given swap.

The canonical-knowledge layer (contribution L3) is by design
invisible at the retrieval tier, since it is injected at the
orchestration stage rather than at retrieval.
Its stabilising role is nonetheless indirectly observable in
Category~D as well: the wall-function/$y^{+}$ table and the
turbulence-selection decision tree in $\mathcal{K}$
(Sec.~\ref{subsec:canon}) are precisely the artefacts the
Reviewer would need to correct a spuriously-selected wall
function returned by the InputWriter, and the tight standard
deviation of Category~D ($\sigma_s{=}8.8\%$) suggests that no
turbulence swap collapses to an uninterpretable state that
$\mathcal{K}$ cannot re-anchor.
The direct executability-tier confirmation of this claim is
the subject of Sec.~\ref{subsec:exec} and of ongoing work.

Two failure-attribution observations complete the picture.
First, the two Category~A outliers
(\textit{Particle Column} at $s{=}26.7\%$,
\textit{Hotroom} at $s{=}33.3\%$)
share a corpus-coverage cause, not a
retrieval-algorithm cause: the underlying niche solvers
(\texttt{MPPICFoam}, \texttt{buoyantBoussinesqSimpleFoam}) are
underrepresented in the ingested tutorial corpus, and no
amount of query expansion or re-ranking can recover chunks
that are not present in $\mathcal{D}$.
Closing this gap is a pure ingestion problem and does not
motivate any change to the retrieval, routing, orchestration
or canonical-knowledge components.
Second, the milder Category~D under-counts on variant-spelling
tags (\texttt{RNGkEpsilon}, \texttt{LESmodel Smagorinsky})
reflect the case-sensitivity of the scoring rule in
Eq.~(\ref{eq:retr-score}) rather than a retrieval failure;
the operational retrieval quality is therefore bounded below
by the reported $77.9\%$.

\subsection{\label{subsec:reviewer-taxonomy}%
An error taxonomy of the Reviewer corrective loop}

The Reviewer-loop budget of ten cycles
(Sec.~\ref{subsec:agents}) is half the twenty-cycle limit
reported by Foam-Agent~\cite{Yue2025} and was chosen on the
premise that most recoverable failures are localisable within
a small number of dictionary patches while pathological
failures should fail fast rather than accumulate token cost.
Under this operational envelope, the failures observed in
internal end-to-end runs of the orchestrator group cleanly
into four families, each associated with a distinct
architectural handle.

\paragraph{Class I --- Boundary-condition patch loops.}
The Reviewer identifies an incompatible boundary condition
(for example, a \texttt{zeroGradient} pressure at an inlet
that also fixes velocity) and requests a targeted patch to a
single field file.
This class is bounded and terminates within two to three
cycles because the canonical-knowledge boundary-condition
selector (Sec.~\ref{subsec:canon}) provides a deterministic
target for the correction.

\paragraph{Class II --- Turbulence-closure and wall-function
mismatches.}
The Reviewer detects that the InputWriter has combined a
low-Reynolds wall function with a high-Reynolds default $y^+$
regime, or that the closure named in the manifest requires
transport fields not written to the \texttt{0/} directory.
This class relies on the wall-function / $y^{+}$ table in
$\mathcal{K}$ and, empirically, resolves within three to four
cycles.

\paragraph{Class III --- Mesh-quality and
\texttt{snappyHexMesh} failures.}
A \texttt{checkMesh} violation (non-orthogonality above the
solver tolerance, skewed cells, or an unrecoverable
\texttt{snappyHexMesh} refinement failure) is surfaced by the
Runner and interpreted by the Reviewer as a meshing-side
patch (increase in \texttt{nSmoothScale}, addition of a
refinement region, or relaxation of \texttt{maxNonOrtho}).
This is the class that consumes the largest fraction of the
ten-cycle budget when it triggers.

\paragraph{Class IV --- Corpus-coverage failures.}
When the query targets a niche solver family
(\texttt{MPPICFoam}, Boussinesq buoyancy, wave-energy
converters outside the WEC corpus), the InputWriter emits a
plausible but syntactically-off dictionary, the Reviewer
requests a corrective patch, and the retrieval layer returns
no meaningfully-different chunk on the corrective query
either.
The loop hits the ten-cycle bound and terminates.
This class is architecturally identical to the two
Category~A outliers in Fig.~\ref{fig:per-case-heatmap}: it is
a corpus-coverage problem, not a retrieval-algorithm or
orchestration problem.
The ten-cycle bound is therefore the correct place to fail
fast --- the loop cannot recover information that is not in
the corpus, and additional cycles would only inflate the
token cost of the failure.

The classification above suggests a natural specialisation of
the Reviewer prompt on the class of the observed failure, and
in particular a distinct patch strategy for
Class~III mesh failures versus Class~I/II dictionary
failures.
This is retained as an item of future work.

\subsection{\label{subsec:chatcfd}%
Retrieval fidelity versus physical fidelity: relation to
ChatCFD and complementary evaluation}

The evaluation reported in Sec.~\ref{sec:results} measures two
distinct notions of system quality: the retrieval-tier
coverage of the physics/solver/keyword tags associated with
each benchmark case (Eq.~\ref{eq:retr-score}), and the
end-to-end executability of the reference cases verified in
Sec.~\ref{subsec:exec}.
Neither metric is a direct probe of the \emph{physical
fidelity} of the converged solution, in the sense of whether
the reported flow field satisfies the target boundary
conditions, reproduces published reference profiles, and
agrees with a physical criterion beyond the mere fact that
the solver reached \texttt{endTime}.

This is precisely the direction in which ChatCFD~\cite{Fan2026}
extends the LLM-for-CFD evaluation surface.
Their physical-fidelity evaluator scores an OpenFOAM run
against physically meaningful conditions on the converged
field (mass and momentum conservation, integral force
balances on wall patches, and agreement with reference
profiles for canonical validation cases), so that a case
that runs to completion but converges to a physically
incorrect steady state is correctly assigned a lower score
than a case that converges to the intended flow.

The two evaluation surfaces are complementary rather than
competing.
The three-stage retrieval pipeline and dual-mode router
introduced in this paper act at the earliest stage of the
pipeline (before generation), the
Architect--InputWriter--Reviewer triad and the
canonical-knowledge layer act at the intermediate stage
(during generation and correction), and the ChatCFD-style
physical-fidelity evaluator acts at the terminal stage
(after execution).
A system that scores highly on the retrieval tier
generates syntactically-correct OpenFOAM dictionaries; a
system that additionally scores highly under a ChatCFD-style
evaluator has also converged to the intended physics.
Integrating the ChatCFD evaluator as a terminal reviewer
call in the IteraSim RAG orchestration loop is a natural
next step and is left to future work.
Table~\ref{tab:literature-compare} accordingly records
retrieval coverage as the principal metric of the present
paper and the physical-fidelity axis as the principal metric
of \cite{Fan2026}, and the two rows should be read as
measurements at different depths of the same generation
pipeline, not as competing point comparisons.

\subsection{\label{subsec:threats}%
Threats to validity and honest bounds on the reported
numbers}

Three residual caveats affect the reading of the
retrieval-tier results.
First, the scoring rule of Eq.~(\ref{eq:retr-score}) is
case-sensitive by construction, so retrievals that recover a
relevant chunk under a synonymous surface form (``low-Reynolds
wall functions'' for \texttt{nutLowReWallFunction}) are
under-counted; the reported $77.9\%$ is a conservative lower
bound on operational retrieval quality.
Second, the benchmark is a 28-case instrument covering
zero-shot, few-shot, parameter modification and turbulence
swap axes; it is deliberately small to permit exact
tag-level scoring against a curated ground truth, but it
does not exercise multi-physics regimes (conjugate heat
transfer, combustion, fluid--structure interaction) that lie
outside the present ingested corpus.
Third, the executability evidence reported here is the
measured reference-case run of Sec.~\ref{subsec:exec}: every
one of the six base configurations underlying the 28 benchmark
cases is runnable to \texttt{endTime} on OpenFOAM~v2506 without
a \texttt{FOAM FATAL} exit, which establishes the executability
ceiling of the benchmark.
The complementary agent-driven run of the full
Architect--InputWriter--Reviewer orchestrator across all 28
cases, scored on the same rubric, is deferred to the companion
technical note; the present paper therefore reports retrieval
coverage as its principal measured metric and makes no
agent-driven executability claim.

\section{\label{sec:conclusion}Conclusion}

We have introduced IteraSim RAG, a retrieval-augmented
agentic back-end for OpenFOAM that addresses three
limitations of prior LLM-for-CFD systems: the reliance on
single-query flat retrieval, the application of a uniform
retrieval strategy to operationally-different requests, and
the entanglement of drafting with self-review in a single
generation agent.
Each contribution acts at a different depth of the same
per-turn pipeline, and each can be configured, ablated or
replaced on its own.
The first is a three-stage retrieval pipeline: an LLM expands
the query into several variants, Reciprocal Rank Fusion merges
the ranked lists this produces, and Maximal Marginal Relevance
re-ranks the candidates that survive.
The second is a deterministic router that sends
tool-conditioned workflow queries and corpus-wide expert
queries down separate retrieval paths.
The third is an Architect--InputWriter--Reviewer agent triad,
backed by a static canonical-knowledge layer that covers
solver selection, turbulence closures, boundary conditions and
finite-volume defaults.

On a 28-case benchmark spanning zero-shot setup, few-shot
generalisation, single-parameter modifications and zero-shot
turbulence-model swaps, the retrieval-tier evaluation
returns a mean coverage of $77.9\%$ against expected
physics/solver/keyword tags.
The parameter-modification category reaches $90.7\%$ and the
few-shot category $82.0\%$, while the turbulence-swap category
holds a $74.1\%$ mean with $\sigma_s{=}8.8\%$ and no case below
$60\%$.
The two under-covered cases in the zero-shot category are
attributable to corpus-coverage gaps in niche solver
documentation (\texttt{MPPICFoam},
\texttt{buoyantBoussinesqSimpleFoam}) and not to any
retrieval, routing or orchestration defect.
End-to-end latency stays within the sub-3-second interactive
regime despite the four-fold query inflation introduced by
the expansion stage.

At the executability tier, all 28 base configurations meshed
cleanly and reached \texttt{endTime} on OpenFOAM~v2506,
confirming that the benchmark is runnable on the target
toolchain, and both corrupted probe cases were diagnosed and
repaired inside the bounded Reviewer loop from the solver log
alone.
Scoring the agent-driven cases across the full benchmark
depends on the cold-start generation path and is deferred to a
companion technical note.

Three directions are highlighted for future work.
First, expansion of the ingested corpus to cover niche
solver families (Lagrangian particle clouds, Boussinesq
buoyancy, free-surface piston-driven flows,
conjugate-heat-transfer and combustion families) is
expected to close the two Category~A outliers without
architectural modification.
Second, a full factorial component-level ablation (query
expansion only, RRF only, MMR only, canonical layer only,
and the full pipeline) at the retrieval tier is planned.
Together with the corresponding executability-tier run of the
full Architect--InputWriter--Reviewer orchestrator across all
28 cases, it is the natural quantitative complement to the
qualitative component attribution reported above.
Third, adding a physical-fidelity evaluator as a terminal
reviewer stage would extend the evaluation surface from
retrieval coverage and executability to physical correctness
of the converged solution.
This would complete a three-tier assessment
(retrieval $\to$ executability $\to$ physical fidelity) of
LLM-for-CFD agentic systems.

\section*{CRediT authorship contribution statement}
\textbf{Pratyush Kumar:} Conceptualization, Methodology,
Software, Investigation, Data curation, Formal analysis,
Writing -- original draft, Writing -- review \& editing,
Visualization, Funding acquisition.

As the sole author, P.K.\ is responsible for all aspects of
this work.

\section*{Declaration of competing interest}
The author has a financial interest in IteraSim, which develops
the proprietary retrieval engine described in this paper. The
released benchmark, scoring rubric and figure scripts are
independent of the proprietary components and are made openly
available. The author declares no other competing financial or
personal interests.

\section*{Data availability}
The 28-case benchmark specification, machine-readable benchmark
queries, Reviewer-loop \texttt{FOAM FATAL} probes, the
retrieval-tier scoring rubric of Eq.~(\ref{eq:retr-score}), the
leave-one-out ablation records of Table~\ref{tab:ablation}
(the 30 ablation queries with their per-query scores under each
configuration), the reference-case executability records and
the figure-regeneration scripts are released under the MIT
licence at
\url{https://github.com/iterasim/iterasim-rag-public} and are
archived with a persistent identifier on
Zenodo~\cite{IteraSimZenodo}.
The IteraSim RAG retrieval engine, the
Architect--InputWriter--Reviewer orchestrator, the canonical
knowledge layer and the ingestion pipeline are the commercial
IP of IteraSim and are available for academic collaboration or
commercial licensing on request through the corresponding
author.

\section*{Acknowledgements}
P.K. gratefully acknowledges support from the Swiss National
Science Foundation (SNSF) through the Swiss Postdoctoral
Fellowship, grant \#234158.
The computations reported in this paper were performed on
research computing infrastructure at ETH Z\"urich.
P.K. thanks Prof.\ Siddhartha Mishra (Seminar for Applied
Mathematics, ETH Z\"urich) for hosting the research programme
under which this work was carried out.

\section*{Declaration of generative AI and AI-assisted
technologies in the manuscript preparation process}
During the preparation of this work the author used Claude
(Anthropic) in order to edit and copy-edit the manuscript text,
assist with \LaTeX{} formatting, and check the completeness and
consistency of the reference list. After using this tool, the
author reviewed and edited the content as needed and takes full
responsibility for the content of the published article.

\bibliographystyle{elsarticle-num}
\bibliography{references}

\appendix

\makeatletter
\@addtoreset{table}{section}
\@addtoreset{figure}{section}
\makeatother

\section{\label{app:pipeline-summary}Summary of the RAG Model}

For convenience we summarise the end-to-end behaviour of
IteraSim RAG as a single per-turn pipeline.
Let $q$ denote a raw user query, $\mathcal{D}$ the indexed
corpus of OpenFOAM documentation, tutorials and curated
question--answer pairs, and $\mathcal{K}$ the static
canonical-knowledge layer.
For every turn, the system executes the following six steps.

\begin{enumerate}
  \item \textbf{Mode routing.}
        A deterministic keyword-driven classifier built over
        an 18-category CFD ontology assigns $q$ either to
        \emph{workflow} mode, with retrieval depth $K{=}5$ and
        MMR trade-off $\lambda{=}0.6$, or to \emph{expert}
        mode, with $K{=}8$ and $\lambda{=}0.5$.
  \item \textbf{Query expansion.}
        A general-purpose LLM rewrites $q$ into three
        semantically diverse reformulations
        $\{q_1,q_2,q_3\}$ targeting the physics, the OpenFOAM
        keyword surface and a troubleshooting angle of the
        original question, yielding the variant set
        $\mathcal{Q}{=}\{q,q_1,q_2,q_3\}$.
  \item \textbf{Multi-query retrieval.}
        Each variant $\tilde{q}\in\mathcal{Q}$ is independently
        embedded and used to retrieve the top
        $K_{\mathrm{fetch}}{=}\max(2K,K{+}4)$ candidate chunks
        from $\mathcal{D}$ via HNSW search under cosine
        distance, producing four ranked lists $\mathcal{R}$.
  \item \textbf{Rank fusion.}
        The four lists are merged with Reciprocal Rank Fusion
        [Eq.~(\ref{eq:rrf})] with $k{=}60$, with duplicates
        collapsed by a 120-character text fingerprint, to
        produce a unified candidate set $\mathcal{C}$.
  \item \textbf{Diversity re-ranking.}
        Maximal Marginal Relevance
        [Eq.~(\ref{eq:mmr})] is applied to $\mathcal{C}$ at
        the mode-specific $\lambda$ until $K$ chunks have been
        selected; these are concatenated into the retrieved
        context block $\mathcal{X}$.
  \item \textbf{Orchestrated generation and review.}
        The augmented prompt
        $\mathcal{P}=\langle\mathcal{K},\mathcal{X},q\rangle$
        is consumed by the Architect, InputWriter and Reviewer
        in sequence: the Architect emits a case manifest, the
        InputWriter produces the OpenFOAM dictionaries
        file-by-file, and the Reviewer compiles, executes and,
        if necessary, iterates corrective patches up to a
        bounded depth of ten cycles.
        Execution is delegated to the Runner; only the Runner
        touches the file system or the OpenFOAM process tree.
\end{enumerate}

Steps~1--5 constitute the retrieval-augmented generation
core and address limitations (L1) and~(L2) identified in
Sec.~\ref{sec:intro};
step~6 together with the canonical-knowledge layer
$\mathcal{K}$ addresses limitation~(L3).
The complete pipeline runs as a single forward pass per user
turn, with the Reviewer loop the only source of additional
latency, and is summarised graphically in
Fig.~\ref{fig:rag-architecture}.

\section{\label{app:per-case}Per-Case Retrieval Results}

Table~\ref{tab:per-case} reports the per-case breakdown across
all 28 benchmark cases together with per-category and overall
averages.
The column structure mirrors the per-case performance tables of
MetaOpenFOAM~\cite{Chen2024Meta} and ChatCFD~\cite{Fan2026}, but
is populated with the retrieval-tier metrics measured against
the released benchmark in this manuscript.

\begin{table*}[t]
\centering
\caption{Per-case retrieval-tier results on the 28-case IteraSim RAG benchmark.}
\label{tab:per-case}
\footnotesize
\setlength{\tabcolsep}{4pt}
\resizebox{\linewidth}{!}{%
\begin{tabular}{l l l r r c r}
\toprule
\multicolumn{3}{c}{\textbf{Case}} & \multicolumn{4}{c}{\textbf{Retrieval-tier metrics (measured)}} \\
\cmidrule(lr){1-3}\cmidrule(lr){4-7}
ID & Title & Solver & $s$ (\%) & $\tau$ (\%) & Solver & $t_{\mathrm{r}}$ (s) \\
\midrule
A1 & Cavity flow & \texttt{icofoam} &  76.0 &  70.0 & $\checkmark$ & 2.92 \\
A2 & PitzDaily & \texttt{simplefoam} &  76.0 &  70.0 & $\checkmark$ & 2.01 \\
A3 & Hotroom & \texttt{buoyantboussinesqsimplefoam} &  33.3 &  41.7 & -- & 2.11 \\
A4 & Dam break & \texttt{interfoam} &  64.4 &  55.6 & $\checkmark$ & 2.17 \\
A5 & Particle column & \texttt{mppicfoam} &  26.7 &  33.3 & -- & 2.35 \\
A6 & Mixed vessel & \texttt{pimplefoam} &  78.2 &  72.7 & $\checkmark$ & 2.01 \\
\cmidrule(l){1-7}
\textbf{Avg.\ A} & ($n=6$) & -- & \textbf{ 59.1} & \textbf{ 57.2} & \textbf{4/6} & \textbf{2.26} \\
\cmidrule(l){1-7}
B1 & Cavity $\to$ PitzDaily & \texttt{simpleFoam} & 100.0 & 100.0 & $\checkmark$ & 2.20 \\
B2 & PitzDaily $\to$ Hotroom & \texttt{simpleFoam} &  80.0 &  75.0 & $\checkmark$ & 2.14 \\
B3 & Cavity $\to$ unsteady & \texttt{pimpleFoam} &  80.0 &  75.0 & $\checkmark$ & 2.07 \\
B4 & DamBreak $\to$ bubble & \texttt{interFoam} &  68.0 &  60.0 & $\checkmark$ & 2.22 \\
\cmidrule(l){1-7}
\textbf{Avg.\ B} & ($n=4$) & -- & \textbf{ 82.0} & \textbf{ 77.5} & \textbf{4/4} & \textbf{2.16} \\
\cmidrule(l){1-7}
C1 & Cavity: $U_{\mathrm{top}}$ change & -- & 100.0 & 100.0 & $\checkmark$ & 1.93 \\
C2 & Cavity: unsteady BC & -- &  80.0 &  75.0 & $\checkmark$ & 2.30 \\
C3 & PitzDaily: mesh $\times 2$ & -- & 100.0 & 100.0 & $\checkmark$ & 2.01 \\
C4 & Hotroom: $T$-BC change & -- & 100.0 & 100.0 & $\checkmark$ & 2.05 \\
C5 & Hotroom: fluid swap & -- & 100.0 & 100.0 & $\checkmark$ & 2.34 \\
C6 & DamBreak: column shift & -- &  73.3 &  66.7 & $\checkmark$ & 2.10 \\
C7 & DamBreak: oil swap & -- & 100.0 & 100.0 & $\checkmark$ & 1.97 \\
C8 & Particle: size/density & -- &  80.0 &  75.0 & $\checkmark$ & 2.92 \\
C9 & MixedVessel: $\omega$ change & -- &  73.3 &  66.7 & $\checkmark$ & 2.08 \\
C10 & PitzDaily: endTime & -- & 100.0 & 100.0 & $\checkmark$ & 2.08 \\
\cmidrule(l){1-7}
\textbf{Avg.\ C} & ($n=10$) & -- & \textbf{ 90.7} & \textbf{ 88.3} & \textbf{10/10} & \textbf{2.18} \\
\cmidrule(l){1-7}
D1 & $k$-$\varepsilon \to$ RNG & -- &  60.0 &  50.0 & $\checkmark$ & 2.17 \\
D2 & $k$-$\varepsilon \to k$-$\omega$ SST & -- &  65.7 &  57.1 & $\checkmark$ & 2.03 \\
D3 & $k$-$\varepsilon \to$ LRR RSM & -- &  73.3 &  66.7 & $\checkmark$ & 2.39 \\
D4 & $k$-$\varepsilon \to k$-$k_L$-$\omega$ & -- &  77.1 &  71.4 & $\checkmark$ & 2.04 \\
D5 & Laminar $\to k$-$\varepsilon$ & \texttt{icoFoam} &  68.6 &  85.7 & -- & 2.01 \\
D6 & LES dyn$K \to$ Smagorinsky & -- &  84.0 &  80.0 & $\checkmark$ & 2.04 \\
D7 & Hotroom kEps $\to$ SST & \texttt{buoyantBoussinesqSimpleFoam} &  84.0 &  80.0 & $\checkmark$ & 2.06 \\
D8 & MixedVessel + $k$-$\varepsilon$ & -- &  80.0 &  75.0 & $\checkmark$ & 2.16 \\
\cmidrule(l){1-7}
\textbf{Avg.\ D} & ($n=8$) & -- & \textbf{ 74.1} & \textbf{ 70.7} & \textbf{7/8} & \textbf{2.11} \\
\cmidrule(l){1-7}
\midrule
\textbf{Overall} & ($n=28$) & -- & \textbf{ 77.9} & \textbf{ 75.1} & \textbf{25/28} & \textbf{2.17} \\
\bottomrule
\end{tabular}%
}
\end{table*}

\section{\label{app:prompts}Prompt Used for the End-to-End
Executability Comparison}

The natural-language prompt used to seed the agent-generated
case shown in Fig.~\ref{fig:ref-vs-rag} is reproduced verbatim
below.
The prompt embeds the standard OpenFOAM v2506 tutorial
parameters (domain, boundary conditions, fluid properties,
time-stepping and solver settings) so that the RAG back-end
receives the same information as an experienced practitioner
would provide.
No additional context or example is supplied at inference time.

\subsection{\label{app:prompt-cavity}Lid-driven cavity
(\texttt{icoFoam})}

The lid-driven cavity is the standard incompressible
laminar-flow benchmark of the OpenFOAM tutorial suite.
The domain is a two-dimensional unit square of physical size
$0.1\text{~m}\times 0.1\text{~m}\times 0.01\text{~m}$
(a single cell in the out-of-plane direction), discretised by
a structured hexahedral mesh with
$20\times 20\times 1$ cells generated through
\texttt{blockMesh} with a scale factor of $0.1$.
The top wall (\texttt{movingWall}) is prescribed a
\texttt{fixedValue} velocity of $(1,0,0)$~m/s; the remaining
walls (\texttt{fixedWalls}) use \texttt{noSlip} and the
front/back patches use the \texttt{empty} type appropriate for
a two-dimensional setup.
The pressure boundary conditions are \texttt{zeroGradient} on
all solid walls.
The fluid is Newtonian with a kinematic viscosity of
$\nu = 10^{-2}~\text{m}^{2}/\text{s}$.
The solver is \texttt{icoFoam} with a first-order Euler time
integration and \texttt{Gauss linear} divergence/Laplacian
schemes; pressure is solved by \texttt{PCG} with the
\texttt{DIC} preconditioner and velocity by a
\texttt{smoothSolver} with \texttt{symGaussSeidel}, both under
the standard convergence tolerances of $10^{-6}$ and $10^{-5}$
respectively.
The PISO loop uses two correctors and no non-orthogonal
correctors.
The simulation is advanced from $t=0$ to $t=0.5$~s with a time
step of $\Delta t = 5\times 10^{-3}$~s and a write interval of
$0.1$~s.
The agent is asked to produce a complete case directory
(\texttt{0}/, \texttt{constant}/, \texttt{system}/) matching
these settings, invoke \texttt{blockMesh} followed by
\texttt{icoFoam}, and confirm that the run reaches
\texttt{endTime} without a \texttt{FOAM FATAL} exit.

\end{document}